\documentclass[
reprint,
superscriptaddress,
amsmath,amssymb,
aps,
]{revtex4-1}

\usepackage{graphicx,epsf,xcolor}

\begin{document}

\title{Electric and thermoelectric properties of graphene bilayers with extrinsic impurities under applied electric field}

\author{G. A. Nemnes}
\email{nemnes@solid.fizica.unibuc.ro}
\affiliation{University of Bucharest, Faculty of Physics, Materials and Devices for Electronics and Optoelectronics Research Center, 077125 Magurele-Ilfov, Romania}
\affiliation{Horia Hulubei National Institute for Physics and Nuclear Engineering, 077126 Magurele-Ilfov, Romania}
\author{T. L. Mitran}
\affiliation{Horia Hulubei National Institute for Physics and Nuclear Engineering, 077126 Magurele-Ilfov, Romania}
\author{A. Manolescu}
\affiliation{School of Science and Engineering, Reykjavik University, Menntavegur 1, IS-101 Reykjavik, Iceland}
\author{Daniela Dragoman}
\affiliation{University of Bucharest, Faculty of Physics, Materials and Devices for Electronics and Optoelectronics Research Center, 077125 Magurele-Ilfov, Romania}
\affiliation{Academy of Romanian Scientists, Splaiul Independentei 54, Bucharest 050094, Romania}

\begin{abstract}
In contrast to monolayer graphene, in bilayer graphene (BLG) one can induce a tunable bandgap by applying an external electric field, which makes it suitable for field effect applications. Extrinsic doping of BLGs enriches the electronic properties of the graphene-based family, as their behavior can be switched from an intrinsic small-gap semiconductor to a degenerate semiconductor. In the framework of density functional theory (DFT) calculations, we investigate the electronic and thermoelectric properties of BLGs doped with extrinsic impurities from groups III (B, Al, Ga), IV (Si, Ge) and V (N, P, As), in the context of applied external electric fields. Doping one monolayer of the BLG with p- or n-type dopants results in a degenerate semiconductor, where the Fermi energy depends on the type of the impurity, but also on the magnitude and orientation of the electric field, which modifies the effective doping concentration. Doping one layer with isoelectronic species like Si and Ge opens a gap, which may be closed upon applying an electric field, in contrast to the pristine BLG. Furthermore, dual doping by III-V elements, in a way that the BLG system is formed by one n-type and one p-type graphene monolayer, leads to intrinsic semiconductor properties with relatively large energy gaps. Si-Si and Ge-Ge substitutions render a metallic like behavior at zero field similar to the standard BLG, however with an asymmetric density of states in the vicinity of the Fermi energy. We analyze the suitability of the highly doped BLG materials for thermoelectric applications, exploiting the large asymmetries of the density of states. In addition, a sign change in the Seebeck coefficient is observed by tuning the electric field as a signature of narrow bands near the Fermi level. 
\end{abstract}

\maketitle

\section{Introduction}

Bilayer graphene (BLG), in comparison to monolayer graphene, is regarded as a more versatile material in the context field effect applications, because one can induce a gap by applying an external electric field \cite{PhysRevB.74.161403,Zhang2009}. Numerous attempts have been considered in order to induce a gap in graphene: flourination \cite{doi:10.1002/smll.201001555,PhysRevB.81.205435,doi:10.1021/nl101437p}, doping with heteroatoms like phosphorus or sulfur \cite{DENIS2013203} even at low doping concentrations \cite{DENIS201720}, by reducing the lateral width of graphene nanoribbons \cite{PhysRevLett.97.216803,Son2006,doi:10.1021/nl0617033}, embedding boron-nitride domains \cite{Liu2013} or the use of a proper substrate \cite{PhysRevLett.106.106801,Zhou2007,PhysRevB.76.073103}. However, extrinsic interventions of this type produce other shortcomings, like reduced carrier mobility. 

On the other hand, the in-plane conduction of intrinsic BLG systems can be controlled simply by applying an electric field perpendicular to the graphene sheets. The presence of the gap is conditioned by breaking the symmetry between A and B type sub-lattices \cite{C4MH00124A}. Because in doped graphenes the gap size is rather small (0.25 eV) and the electric field required is large, the use of intrinsic BLG is less practical, in spite of the gap tunability. Therefore, in pursuing the gap enhancement in BLGs, similar to the case of monolayer graphene, several methods have been considered, such as doping \cite{DENIS2010251,doi:10.1002/cphc.201402608,DENIS2016152,PhysRevB.82.245414,Fujimoto2015,NEMNES2018175}, functionalization \cite{PhysRevB.78.085413,HU201475} or substrate influence \cite{Ohta951,C7RA01134B}.   

Chemical doping of BLGs was further explored from different perspectives. The junction formed between metallic contacts and graphene as well as doping with metallic species like Au revealed that a high density of electrons can be transferred from the upper donor layer to the lower one \cite{1367-2630-12-3-033046}. 
Manganese doped BLGs present a highly polarized spin state suitable for spintronic applications \cite{0957-4484-19-20-205708}, similar to single layer graphene nanoribbons \cite{NEMNES1,NEMNES20131347}. Using Mn doping a ferromagnetic graphene field-effect transistor with a finite band gap was fabricated \cite{C5TC00051C}. Molecular doping is another option which provides the emergence of permanent electronic and optical band gaps in BLGs upon adsorption of $\pi$-electron molecules \cite{doi:10.1021/nn400340q}. Furthermore, investigations concerning the differences between molecular doping with electron transfer molecules and gating  revealed important insights for molecular electronics applications \cite{Uchiyama2017}. More recently K-doped BLGs, with intercalated impurities, were shown to exhibit n-type conduction \cite{doi:10.1063/1.5012808}, just alike similarly doped graphene nanotubes.

Single and dual doping has been extensively investigated in mono- and bilayer graphene systems \cite{DENIS2016152}. In this paper we are particularly focused on the combined effect of doping and external electric field. We already showed that the effective doping by boron or nitrogen can be systematically modified by an applied electric field \cite{NEMNES2018175}. It was also established that dual doping by B and N, opens a gap at zero field in a BLG system. Here we investigate comparatively these effects by considering other elements from group III (B, Al, Ga), group IV (Si, Ge), group V (N, P, As) and also dual doping (III-V, IV-IV). In the latter case, we opted for a BLG system with one n-type monolayer and one p-type monolayer.

Controlling the p- and n-type doping of BLG would further support the idea of achieving a CMOS type technology on a single bilayer system. Logic circuits depend on the ability of integrating normally-ON and normally-OFF FETs. Inducing a gap in BLG systems without applying an electric field, which can be closed when a top gate voltage is applied would complement the behavior of the pristine BLG. 
Degenerate semiconductors may find applications as contact regions or transparent electrodes, but can also serve in other opto-electronic and thermoelectric applications. From the former category one can mention highly doped p-n type devices such as tunneling diodes, while from the latter, cooling nanodevices and thermal energy harvesters.   
A tunable Fermi energy and density of states (DOS) by an external electric field represents here an asset. 
We approach these issues in a systematic investigation of the electronic and thermolectric properties of doped BLG systems.

\section{Model systems and computational methods}

\begin{figure}[t]
\centering
\includegraphics[width=7cm]{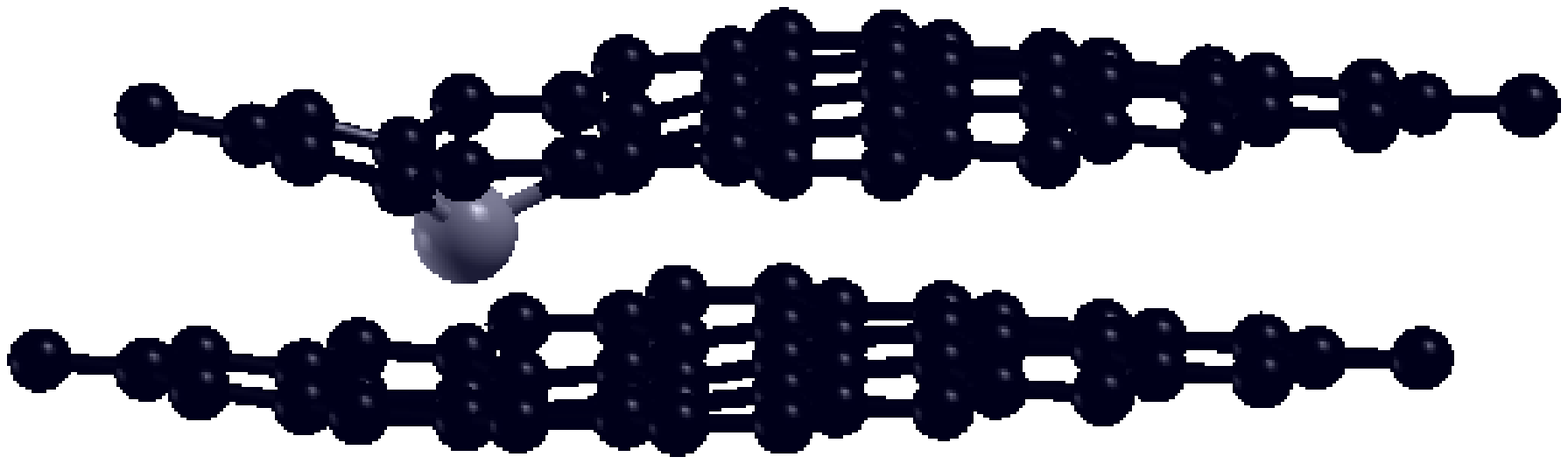} (a)\\
\vspace*{0.8cm}\includegraphics[width=7cm]{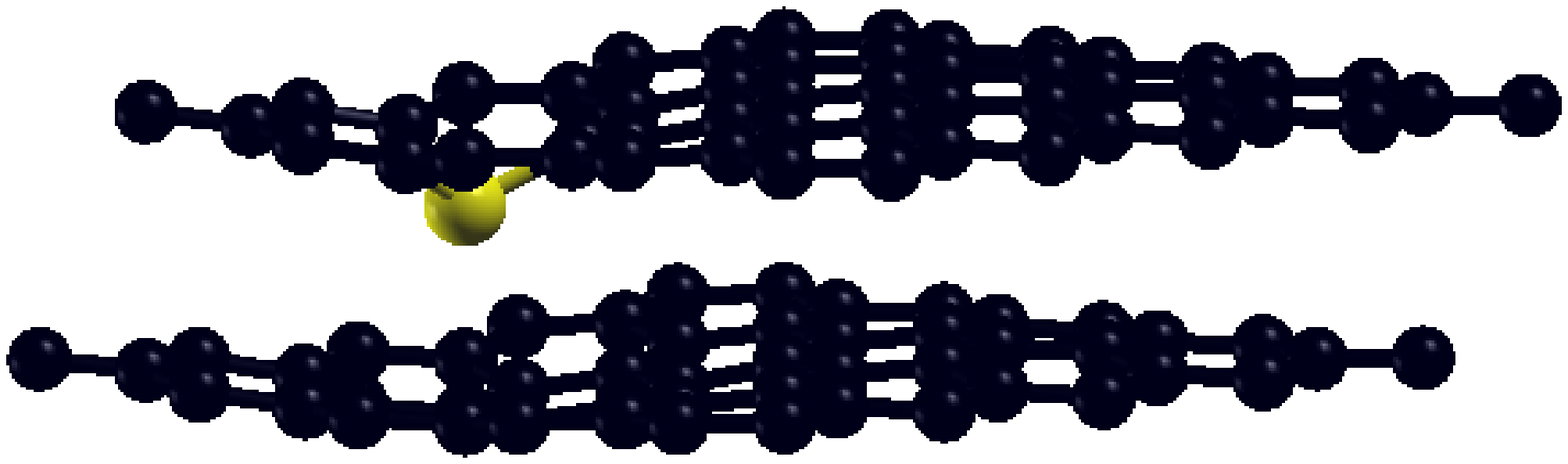} (b)\\
\vspace*{0.8cm}\includegraphics[width=7cm]{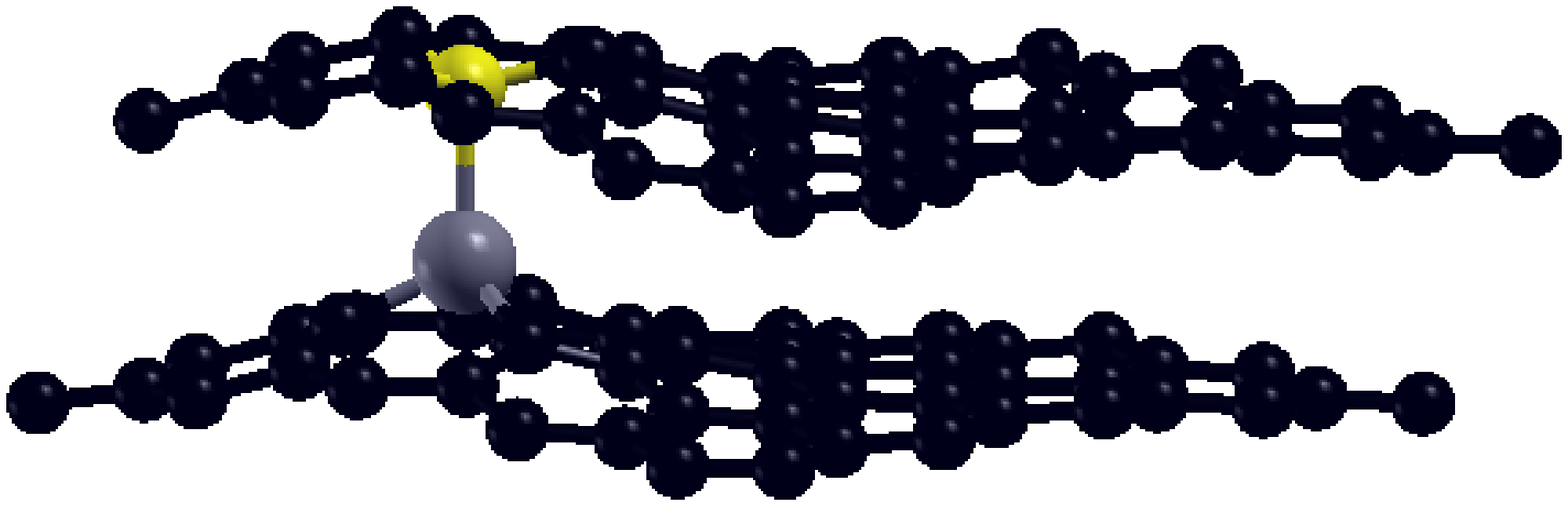} (c)
\caption{Typical structural conformations of doped BLG structures: (a) p-type doping by Al; (b) n-type doping by P; (c) dual doping by Al and P.}
\label{structures}
\end{figure}

The BLG systems investigated here have the standard $AB$ type stacking, which allows the symmetry breaking between the two layers when an electric field is applied. We denote by $A_s$ and $B_s$ the two sub-lattices in the graphene sheets, where $s=1,2$ is the index of the monolayer. Given the $AB$ stacking of the BLG, we choose that $B_1$ and $B_2$ sites overlap, while $A_1$ and $A_2$ correspond to the middle of a hexagon in the other layer. We analyze $5\times5$ structures with 100 atoms in the supercell with one or two impurities from group III (B, Al, Ga), group IV (Si, Ge), group V (N, P, As). For single impurity doping the site positions $A_1$ and $B_1$ were considered, while for dual doping we take $B_1$ and $B_2$ with the maximum overlap. Typical structural conformations are indicated in Fig.\ \ref{structures}. In the following we label the structures with one substitutional impurity as $X$-BLG and the structures with two impurities as $XY$-BLG.

The {\it ab initio} DFT calculations are performed using the SIESTA package \cite{0953-8984-14-11-302}, which achieves a linear scaling of the computational time with the system size by employing a strictly localized basis set. The double-$\zeta$ (DZ) basis set was used, with a real space grid set by a 500 Ry grid cuttoff to remove any potential egg-box effects. Typical LDA or GGA functionals underestimate the interlayer distance of 3.35 \AA\ determined experimentally for the pristine BLG. Since this is a particularly important aspect for the electronic structures of doped BLGs, we employ here the more computationally expensive vdW-DF functional of Dion {\it et al.} (DRSLL) \cite{PhysRevLett.92.246401}, which is able to reproduce the interlayer distance and electronic properties of pristine BLG with high accuracy \cite{Birowska_Milowska_Majewski_2011}. The pseudopotentials of Troullier and Martins \cite{PhysRevB.43.1993} were used, with a typical valence electron configurations. A k-point sampling scheme of $11\times11\times1$ was employed for the integrals in the 1BZ. However, the density of states (DOS) was generated on a finer grid of $101\times101\times1$ k-points. Similar to Refs.\ \cite{Zhang2009,PhysRevLett.115.015502} the electronic properties of doped BLGs are analyzed considering external electric fields, applied perpendicular to the graphene sheets. 
 
\section{Results and discussion}

\subsection{Structural properties}

Following structural relaxations, local deformations due to the substitutional impurities {placed in $B_1$ and $B_1-B_2$ positions} appear in the BLG, as indicated in Figs.\ S1 and S2, which are summarized Table \ref{tab1}. 
\begin{table}[h]
\centering
\caption{Structural properties of BLGs with substitutional impurities from  groups III, IV and V {in $B_1$ position}: average bond length to neighboring C atoms $d_{X-C}$, off-plane shift $\Delta z_X$, BLG inter-layer distance $d_{\rm BLG}$. All distances are given in \AA.}
\scriptsize
\vspace*{0.3cm}
\begin{tabular}{c|ccc|cc|ccc}
\hline\hline
d [\AA]              & B    & Al   & Ga   & Si   & Ge   & N    & P    & As  \\
\hline
$d_{X-C}$      & 1.47 & 1.86 & 1.94 & 1.74 & 1.82 & 1.41 & 1.77 & 1.88 \\ 
$\Delta z_X$  & 0.08 & 1.73 & 1.79 & 1.39 & 1.58 & 0.05 & 1.36 & 1.51 \\
$d_{\rm BLG}$ & 3.37 & 3.45 & 3.47 & 3.45 & 3.48 & 3.36 & 3.43 & 3.44 \\
\hline
\hline
\end{tabular}
\label{tab1}
\centering
\caption{Structural properties of BLGs with dual doping with III-V elements in $B_1-B_2$ positions: $d_{X-C}$, $d_{Y-C}$, $d_{X-Y}$, $\Delta z_X$, $\Delta z_Y$ and $d_{\rm BLG}$. All distances are given in \AA.}
\scriptsize
\vspace*{0.3cm}
\begin{tabular}{c|ccc|ccc|ccc}
\hline\hline
d [\AA] & B-N    & B-P   & B-As   & Al-N   & Al-P   & Al-As    & Ga-N    & Ga-P & Ga-As \\
\hline
$d_{X-C}$   & 1.47 & 1.50 & 1.51 & 1.84 & 1.85 & 1.88 & 1.91 & 1.95 & 1.96 \\ 
$d_{Y-C}$   & 1.41 & 1.76 & 1.86 & 1.45 & 1.72 & 1.92 & 1.44 & 1.72 & 1.91 \\
$d_{X-Y}$   & 3.26 & 2.66 & 2.65 & 2.14 & 2.35 & 3.54 & 2.32 & 2.40 & 2.41 \\ 
$\Delta z_X$  & 0.07 & 0.79 & 0.95 & 1.89 & 1.92 & 2.17 & 1.98 & 2.16 & 2.19 \\
$\Delta z_Y$  & 0.04 & 1.49 & 1.68 & 0.52 & 0.75 & 1.94 & 0.72 & 0.99 & 1.03 \\
$d_{\rm BLG}$ & 3.37 & 3.37 & 3.39 & 3.51 & 3.53 & 3.31 & 3.58 & 3.58 & 3.57 \\
\hline
\hline
\end{tabular}
\label{tab2}
\end{table}

For the $X$-BLGs we determined the average bond length between the impurity $X$ and neighboring C atoms $d_{X-C}$, the resulting shift from the graphene plane $\Delta z_X$ and the modifications to the BLG inter-layer distance $d_{\rm BLG}$. For all three groups of elements there is a systematic increase of the distances with the atomic number. Group-V elements introduce smaller deformations compared to the corresponding group-III elements. It worth noting that $d_{\rm BLG}$ increases in all cases. A similar behavior is found for $A_1$ positions as indicated in Table S1.

For dual doping by $X$-$Y$ pairs, the structural data is indicated in Table\ \ref{tab2}, where $X$ = B, Al, Ga and $Y$ = N, P, As. The values are in general close to the ones obtained for single impurity doping. Al and Ga combined with group V elements increase significantly the BLG inter-layer distance up to 3.58 \AA. In addition, the distance between the two impurities $d_{X-Y}$ is typically much shorter than $d_{\rm BLG}$, one exception being the B-N pair, where the off-plane shifts are negligible. For low dopant concentrations, it is expected that the dopants would interact with carbon atoms from the other layer and the local structural configuration of the system would be much alike the case with one impurity, while the interlayer distance would be less perturbed. Furthermore, as pointed out in Ref. \cite{C6CP02481E} it is possible that the dopants protrude outside of the bilayer system, in which case the interlayer distance becomes smaller, comparable to the pristine BLG.

\subsection{Electronic properties}

\begin{figure}[t]
\centering
\includegraphics[width=\columnwidth]{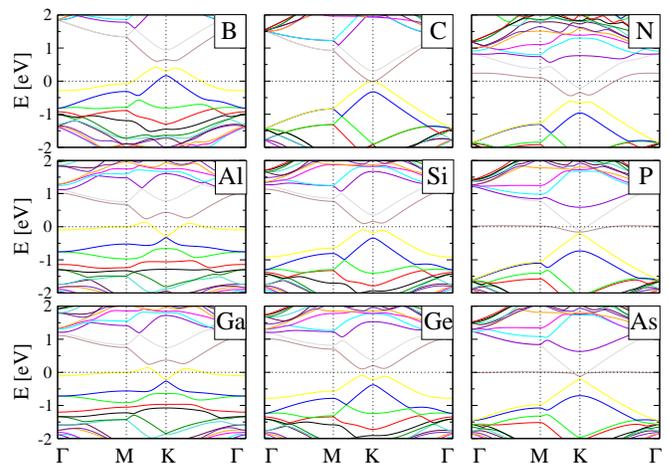}
\caption{Electronic band structures of doped $5\times5$ BLGs with one impurity in $B_1$ position for zero electric field. The impurities correspond to group III (B, Al, Ga), group IV (C -- meaning no impurity, Si, Ge) and group V (N, P, As) elements. The Fermi level corresponds to $E=0$ eV.}
\label{bands-1imp}
\end{figure}

\begin{figure}[t]
\centering
\includegraphics[width=\columnwidth]{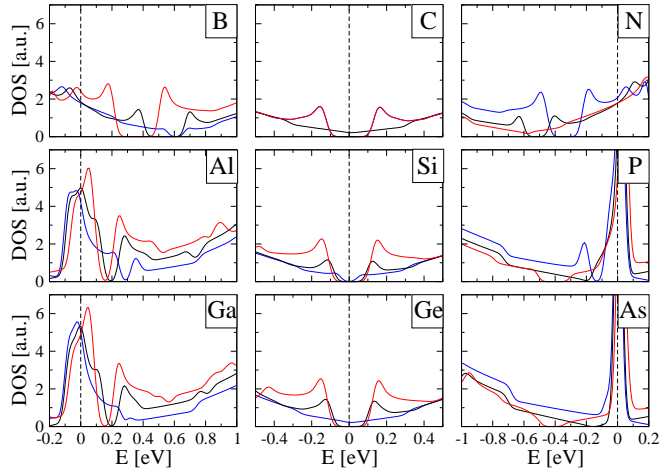}
\caption{The density of states for the doped BLG systems indicated in Fig.\ \ref{bands-1imp}. In each plot there are three values for the electric field: $E_{\rm field} =$ -0.5 V/nm (blue), 0 V/nm (black) and 0.5 V/nm (red). The Fermi level is marked by vertical dashed lines. In the unperturbed case (C as "impuritiy") the blue and black curves overlap.}
\label{DOS-1imp}
\end{figure}

We first investigate BLG systems doped with one impurity in the upper graphene monolayer, in position $B_1$. The band structures determined for zero electric field are indicated in Fig.\ \ref{bands-1imp}. Due to the relatively high doping, with one impurity for 100 C atoms in the supercell, group-III and group-V doping render a degenerate semiconductor with the anticipated behavior: for B, Al, Ga doping the system becomes p-type, while for N, P, As doping the system is n-type. Still, there are some distinctive features within systems with impurities from the same group. Increasing the atomic number $Z_{X}$, one band presenting a rather small dispersion is located near the Fermi level. The energy gap, which is absent in the pristine BLG, is located above or below the Fermi level, depending on the extrinsic impurity type. The charge redistribution is equivalent to an internal electric field. One observes a systematic behavior regarding the position of the gap, i.e. it is getting closer to the Fermi level, as $Z_{X}$ increases. Consequently, the Fermi energy of the degenerate semiconductor, defined as the difference between the Fermi level and the bottom of the conduction band / top of the valence band, for n- and p- type semiconductors, respectively, decreases along with $Z_{X}$. On the other hand, isoelectronic extrinsic impurities introduce small energy gaps, revealing an intrinsic semiconductor behavior. In Fig.\ \ref{bands-1imp} we also show the band structure of the unperturbed BLG, when the "impurities" are C atoms themselves, compared to Si and Ge impurities from group IV. The effect of the Si and Ge impurities is to open the gap at the Fermi energy.

The density of states (DOS) of the single doped BLG systems are shown in Fig.\ \ref{DOS-1imp}, for zero electric field and an electric field $|\vec{E}_{\rm field}|=5$ V/nm perpendicular to the graphene sheets, with the two possible orientations. The position of the energy gap is systematically modified by the external electric field, as already indicated for B and N doped BLGs \cite{NEMNES2018175}. The gaps introduced by either B or N substitutions were also found in the context of disorder using the random tight-binding model \cite{MOUSAVI201890}. A particular feature is the appearance of pronounced peaks in the DOS for Al-BLG and Ga-BLG systems located at the Fermi level, which become sharper for P-BLG and As-BLG. The peaks correspond to the poorly dispersive bands indicated in Fig.\ \ref{bands-1imp}, which were also observed in Ref. \cite{DENIS2010251}. As it is shown later, the sharp variation of the DOS, which may be further influenced by the applied electric field has significant consequences regarding the magnitude and sign of the Seebeck coefficient. Overall, the features of the electronic structure are well reproduced also for $A_1$ positions of the substitutional impurities, as one may see from Figs. S3 and S4 in the SM. To establish which configuration is more probable, we calculated the formation energies $E_{\rm f}$ for both sub-lattice positions, $A_1$ and $B_1$, using the relation $E_{\rm f} = E_{X-\rm BLG} - E_{\rm BLG} + E_{\rm C} - E_X$, where  $E_{X-\rm BLG}$, $E_{\rm BLG}$, $E_{\rm C}$ and $E_X$ are the total energies of the doped BLG, pristine BLG and isolated C and X atoms, respectively \cite{DENIS2010251}. The values are listed in Table S1, showing that, generally, $B_1$ substitutions are energetically slightly favored over $A_1$ substitutions. Concerning Si-BLG and Ge-BLG, in the case of $B_1$ substitutions, the gap observed at zero electric field can be closed by an external field with proper orientation, in contrast to the pristine BLG. Based on this observation, circuits implementing binary logic can be constructed as in standard CMOS, where high/low potential on the top gate can switch the conduction between ON and OFF, but also the other way around, i.e. in normally-ON and normally-OFF transistors. 

\begin{figure}[t]
\centering
\includegraphics[width=\columnwidth]{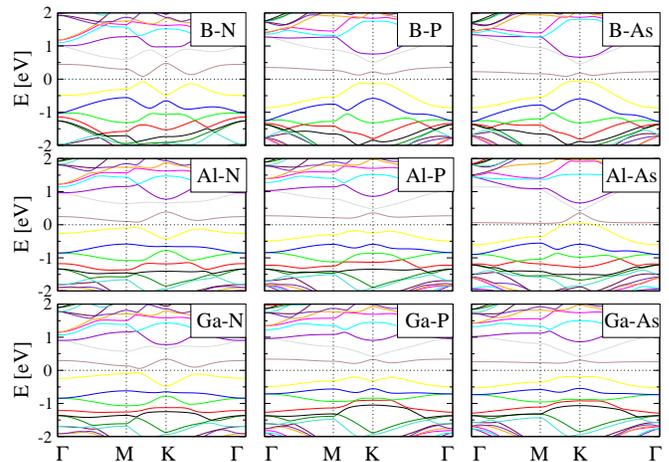}
\caption{Electronic band structures of dual doped BLGs by III-V elements. Intrinsic semiconductor behavior is found, except Al-As and Ga-As which introduce larger structural deformations in the BLG. }
\label{bands-2imps}
\end{figure}

\begin{figure}[t]
\centering
\includegraphics[width=\columnwidth]{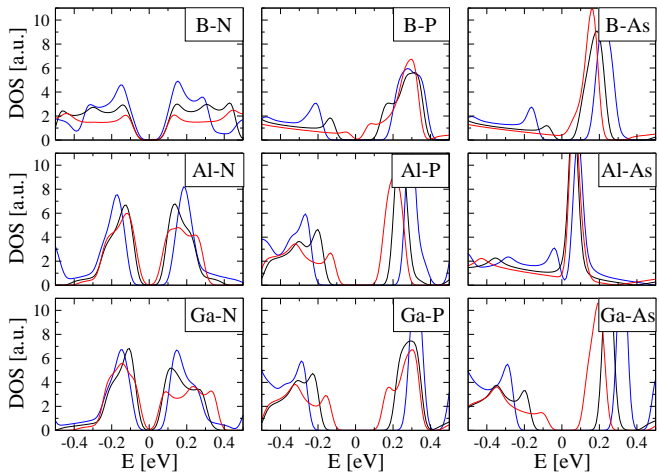}
\caption{The density of states of dual doped BLGs by III-V elements. Particularly large energy gaps are found for Al-P, Ga-P, Ga-As in the absence of the external electric field, which are further enhanced at negative fields ($E_{\rm field}=-5$ V/nm).}
\label{DOS-2imps}
\end{figure}

Next, we turn our attention to intrinsic BLG systems obtained by dual doping with group-III and group-V elements. For the subsequent analysis we chose $B_1$-$B_2$ sites to maximize the charge transfer between the two impurities, but the general behavior is not strongly influenced by the sub-lattice positions. With one p-type impurity in the lower layer and one n-type impurity in the upper layer, the BLG system becomes an intrinsic semiconductor, as one can see from the band structures presented in Fig.\ \ref{bands-2imps}. {As it was shown in previous studies \cite{MOUSAVI201890,NEMNES2018175}, B-N dual doping produces small gaps, symmetric with respect to the Fermi level, which are present irrespective of the sub-lattice positions.} A negatively oriented electric field ($E_{\rm field}=-5$ V/nm) enhances the local field produced by the impurity pair and the gaps are getting larger, while the opposite is found for positive fields ($E_{\rm field}= 5$ V/nm). Although for BN-BLG system there is a relatively low tunability of the gap, it becomes progressively larger as heavier impurities are involved, either p-type or n-type, as one can see from Fig.\ \ref{DOS-2imps}. BP-BLG and BAs-BLG have significant gaps, which in addition show a good tunability with the electric field. The largest gap of $\sim$ 0.4 eV ($\sim$ 0.45 eV with applied field) is found for GaAs-BLG. Other dual doped BLGs have been predicted to exhibit large gaps \cite{DENIS2016152}, but the $X$-$Y$ substitutions were located in one of the two graphene monolayers. From the fabrication point of view it is however easier to use graphene monolayers with a certain type of doping and subsequently be assembled into BLGs. For comparison, dual doping with isoelectronic species like Si or Ge results in a finite, but asymmetric DOS when no electric field is applied. The SiSi-BLG and GeGe-BLG systems, with the impurities placed on $B_1$-$B_2$ positions, obey the reflection symmetry $z \rightarrow -z$ such that the two orientations of $E_{\rm field}$ are equivalent. In this case, a small gap appears, as one can see in Fig. S5 in the SM. This behavior is quite similar to the one of pristine BLG, apart from the pronounced asymmetry in the DOS.

\subsection{Seebeck coefficients}

In the context of electric field tuning of the DOS, we investigate the behavior of the Seebeck coefficient for the BLG systems with single impurity type doping, which indicate degenerate semiconductor behavior.
In general, the electric conductivity $\sigma$ and the Seebeck coefficient $S$ can be calculated using the linear response functions $L_n$ ($n = 0, 1, 2, \ldots$):
\begin{equation}
L_n = \int dE \sigma(E) \left(\frac{\partial f_{\rm FD}}{\partial E}\right) (E-E_{\rm F})^n, 
\end{equation} 
where 
\begin{eqnarray} 
\sigma(E) &=& \int dE' D(E') \frac{v_g^2(E')}{d} \tau(E') \delta(E-E') \nonumber \\  
          &=& D(E) \frac{v_g^2(E)}{d} \tau(E),
\end{eqnarray}
$D(E)$ is the total DOS, $f_{\rm FD}$ is the Fermi-Dirac distribution, $v_g=1/\hbar({\partial E}/{\partial k})$ is the group velocity, $\tau$ is the relaxation time and $d=2$ is the system dimensionality.
For a given temperature $T$, the electrical conductivity $\sigma(T)$ and thermopower $S(T)$ are given by:
\begin{eqnarray}
\sigma(T) &=& e^2 L_0, \\ 
S(T) &=& -\frac{1}{e T} \frac{L_1}{L_0}.
\end{eqnarray}
For metals or degenerate semiconductors and low temperatures, the expression of the Seebeck coefficient takes the Cutler-Mott form \cite{PhysRev.181.1336}:
\begin{equation}
S_{\rm CM} = -\frac{\pi^2 k_{\rm B}T}{3 e} \left. \frac{d [\ln \sigma(E)]}{dE}\right|_{E=E_{\rm F}}
  = -\frac{\pi^2 k_{\rm B}T}{3 e \sigma} \left. \frac{d [\sigma(E)]}{dE}\right|_{E=E_{\rm F}}.
\end{equation}
Furthermore, assuming that near the Fermi level the relaxation time and the group velocity do not depend on the energy, the expression of the Seebeck coefficient $S_{\rm CM}$ simplifies considerably and it may be written using D(E):
\begin{equation} 
\label{S-CM-DOS}
S_{\rm CM} = -\frac{\pi^2 k_{\rm B}T}{3 e D(E_{\rm F})} \left. \frac{d [D(E)]}{dE}\right|_{E=E_{\rm F}}.
\end{equation}
From Eq.\ (\ref{S-CM-DOS}) one notes that $S_{\rm CM}$ is maximized for large asymmetries in the DOS in the vicinity of $E_{\rm F}$ and it is also proportional with $T$ at low temperatures. However, in experimentally relevant conditions, the physical properties, like the electrical conductivity may depend on a number of factors, which follow from the preparation conditions, such as system crystallinity and point defects.

In the previous section we showed that the group-III and group-V doped BLGs exhibit degenerate semiconductor characteristics. Moreover, they have particular features in the DOS, which render qualitatively different regimes. We analyze in the following the low temperature behavior of the Seebeck coefficient, in the Cutler-Mott approximation. For this purpose, we denote by $\alpha_{\rm CM}$ the proportionality constant in the linear dependence of $S_{\rm CM}(T)$ in Eq.\ (\ref{S-CM-DOS}), i.e. $S_{\rm CM} = \alpha_{\rm CM} T$ and the calculated values of $\alpha_{\rm CM}$ for $E_{\rm field} = 0, \pm 5$ V/nm are listed in Table\ \ref{tab3} for $B_1$ type substitutions.

\begin{table}[h]
\centering
\caption{Thermoelectric behavior at low temperatures of doped BLG systems under applied electric field: $\alpha_{\rm CM} = S_{\rm CM} / T $ in units of 10$^{-2}$ $\mu$V/K$^2$.}
\scriptsize
\vspace*{0.3cm}
\begin{tabular}{c|ccc|ccc}
\hline\hline
$E_{\rm field}$ [V/nm]  & B    & Al   & Ga   & N    & P    & As  \\
\hline
-5 V/nm      & 5.62 & 18.61 &  12.61 & -10.15 & -29.68 & -10.23  \\ 
 0           & 9.94 &  1.42 &  -0.13 &  -8.04 & -50.05 & -17.61  \\
+5 V/nm      & 7.68 & -9.60 & -13.06 &  -5.55 & -66.97 & -73.71  \\
\hline
\hline
\end{tabular}
\label{tab3}
\end{table}

For $[dD(E)/dE]|_{\rm E_F}>0$ the Seebeck coefficient is negative, which corresponds to the n-type conduction.
This is a typical case for N, P, As doped systems. As one can see from Fig.\ \ref{DOS-1imp} the DOS is increasing with energy at $E_{\rm F}$, while for P-BLG and As-BLG systems the sharp peak in the DOS is centered at slightly higher energies than $E_{\rm F}$. The opposite situation, $[dD(E)/dE]|_{\rm E_F}>0$ corresponds to p-type systems. From Table\ \ref{tab3} we notice that B and Al doped systems at zero electric field have positive Seebeck coefficients, while the Ga doped BLG has a small negative value. This corresponds to a decrease in the DOS for B-doped BLG, while for Al and Ga doped BLGs a broad peak is located around the Fermi level. Applying an external electric field, the n- or p-type behavior found in the absence of the field is retained. A distinct behavior is found for Al-BLG and Ga-BLG, where the sign of the Seebeck coefficient changes, as $E_{\rm field}$ is tuned from negative into positive values and the peak located near the Fermi is shifted. Similar trends are also obtained for the $A_1$ position of the subtitutional impurities, as shown in Table S2. A behavior of this type has been reported in low dimensional systems \cite{NEMNES20101613,1367-2630-14-3-033041,PhysRevLett.119.036804}, in the context of resonant transport, where the position of resonances may be tuned during the operation conditions, whereas here we predict it in a bulk system. Furthermore, band engineering is considered a potential route for achieving a tunable sign of the thermopower even if either n- or p-type doping cannot be obtained \cite{PhysRevLett.112.196603}. If this can be realized dynamically by an external stimulus, e.g. by applying an external electric field, it may serve for reconfiguring Peltier devices at the nanoscale.



\section{Conclusions}

We investigated the effect of extrinsic doping on the electric and thermoelectric properties of BLGs, in the context of applied electric fields. Group-III and group-V impurities render the system p-type and n-type, respectively, and under the conditions of high doping the systems behave as degenerate semiconductors with a tunable Fermi energy by the external electric field. This may provide applications to tunneling diodes, formed either in-plane or in stacked layers. Si and Ge substitutions induce a gap at zero field, which may be closed by applying a properly oriented electric field, which is the opposite behavior of the pristine BLG. This is particularly important for achieving a complementary behavior with respect to pristine BLG, so that normally-ON and normally-OFF field effect transistors can be embedded in the same BLG, resembling the CMOS circuitry design. Furthermore, the asymmetries of the DOS near the Fermi level, which can be modified by the external field, can be exploited in thermoelectric devices. In particular, Al and Ga doped systems may exhibit a sign change in the Seebeck coefficient by tuning $E_{\rm field}$. By dual doping with III-V elements, a semiconductor behavior is obtained, with significant gaps of $\sim$ 0.45 eV for Ga-As.
These observations enrich the potential applications of BLGs, from integrated logic circuits to cooling nano-elements.    \\


\acknowledgments
This work was supported by the Romanian Ministry of Research and Innovation under the project PN 19060205/2019.


\bibliography{manuscript_R1}

\begin{thebibliography}{45}%
\makeatletter
\providecommand \@ifxundefined [1]{%
 \@ifx{#1\undefined}
}%
\providecommand \@ifnum [1]{%
 \ifnum #1\expandafter \@firstoftwo
 \else \expandafter \@secondoftwo
 \fi
}%
\providecommand \@ifx [1]{%
 \ifx #1\expandafter \@firstoftwo
 \else \expandafter \@secondoftwo
 \fi
}%
\providecommand \natexlab [1]{#1}%
\providecommand \enquote  [1]{``#1''}%
\providecommand \bibnamefont  [1]{#1}%
\providecommand \bibfnamefont [1]{#1}%
\providecommand \citenamefont [1]{#1}%
\providecommand \href@noop [0]{\@secondoftwo}%
\providecommand \href [0]{\begingroup \@sanitize@url \@href}%
\providecommand \@href[1]{\@@startlink{#1}\@@href}%
\providecommand \@@href[1]{\endgroup#1\@@endlink}%
\providecommand \@sanitize@url [0]{\catcode `\\12\catcode `\$12\catcode
  `\&12\catcode `\#12\catcode `\^12\catcode `\_12\catcode `\%12\relax}%
\providecommand \@@startlink[1]{}%
\providecommand \@@endlink[0]{}%
\providecommand \url  [0]{\begingroup\@sanitize@url \@url }%
\providecommand \@url [1]{\endgroup\@href {#1}{\urlprefix }}%
\providecommand \urlprefix  [0]{URL }%
\providecommand \Eprint [0]{\href }%
\providecommand \doibase [0]{http://dx.doi.org/}%
\providecommand \selectlanguage [0]{\@gobble}%
\providecommand \bibinfo  [0]{\@secondoftwo}%
\providecommand \bibfield  [0]{\@secondoftwo}%
\providecommand \translation [1]{[#1]}%
\providecommand \BibitemOpen [0]{}%
\providecommand \bibitemStop [0]{}%
\providecommand \bibitemNoStop [0]{.\EOS\space}%
\providecommand \EOS [0]{\spacefactor3000\relax}%
\providecommand \BibitemShut  [1]{\csname bibitem#1\endcsname}%
\let\auto@bib@innerbib\@empty
\bibitem [{\citenamefont {McCann}(2006)}]{PhysRevB.74.161403}%
  \BibitemOpen
  \bibfield  {author} {\bibinfo {author} {\bibfnamefont {E.}~\bibnamefont
  {McCann}},\ }\href@noop {} {\bibfield  {journal} {\bibinfo  {journal} {Phys.
  Rev. B}\ }\textbf {\bibinfo {volume} {74}},\ \bibinfo {pages} {161403}
  (\bibinfo {year} {2006})}\BibitemShut {NoStop}%
\bibitem [{\citenamefont {Zhang}\ \emph {et~al.}(2009)\citenamefont {Zhang},
  \citenamefont {Tang}, \citenamefont {Girit}, \citenamefont {Hao},
  \citenamefont {Martin}, \citenamefont {Zettl}, \citenamefont {Crommie},
  \citenamefont {Shen},\ and\ \citenamefont {Wang}}]{Zhang2009}%
  \BibitemOpen
  \bibfield  {author} {\bibinfo {author} {\bibfnamefont {Y.}~\bibnamefont
  {Zhang}}, \bibinfo {author} {\bibfnamefont {T.-T.}\ \bibnamefont {Tang}},
  \bibinfo {author} {\bibfnamefont {C.}~\bibnamefont {Girit}}, \bibinfo
  {author} {\bibfnamefont {Z.}~\bibnamefont {Hao}}, \bibinfo {author}
  {\bibfnamefont {M.~C.}\ \bibnamefont {Martin}}, \bibinfo {author}
  {\bibfnamefont {A.}~\bibnamefont {Zettl}}, \bibinfo {author} {\bibfnamefont
  {M.~F.}\ \bibnamefont {Crommie}}, \bibinfo {author} {\bibfnamefont {Y.~R.}\
  \bibnamefont {Shen}}, \ and\ \bibinfo {author} {\bibfnamefont
  {F.}~\bibnamefont {Wang}},\ }\href@noop {} {\bibfield  {journal} {\bibinfo
  {journal} {Nature}\ }\textbf {\bibinfo {volume} {459}},\ \bibinfo {pages}
  {820} (\bibinfo {year} {2009})}\BibitemShut {NoStop}%
\bibitem [{\citenamefont {Nair}\ \emph {et~al.}(2010)\citenamefont {Nair},
  \citenamefont {Ren}, \citenamefont {Jalil}, \citenamefont {Riaz},
  \citenamefont {Kravets}, \citenamefont {Britnell}, \citenamefont {Blake},
  \citenamefont {Schedin}, \citenamefont {Mayorov}, \citenamefont {Yuan},
  \citenamefont {Katsnelson}, \citenamefont {Cheng}, \citenamefont
  {Strupinski}, \citenamefont {Bulusheva}, \citenamefont {Okotrub},
  \citenamefont {Grigorieva}, \citenamefont {Grigorenko}, \citenamefont
  {Novoselov},\ and\ \citenamefont {Geim}}]{doi:10.1002/smll.201001555}%
  \BibitemOpen
  \bibfield  {author} {\bibinfo {author} {\bibfnamefont {R.~R.}\ \bibnamefont
  {Nair}}, \bibinfo {author} {\bibfnamefont {W.}~\bibnamefont {Ren}}, \bibinfo
  {author} {\bibfnamefont {R.}~\bibnamefont {Jalil}}, \bibinfo {author}
  {\bibfnamefont {I.}~\bibnamefont {Riaz}}, \bibinfo {author} {\bibfnamefont
  {V.~G.}\ \bibnamefont {Kravets}}, \bibinfo {author} {\bibfnamefont
  {L.}~\bibnamefont {Britnell}}, \bibinfo {author} {\bibfnamefont
  {P.}~\bibnamefont {Blake}}, \bibinfo {author} {\bibfnamefont
  {F.}~\bibnamefont {Schedin}}, \bibinfo {author} {\bibfnamefont {A.~S.}\
  \bibnamefont {Mayorov}}, \bibinfo {author} {\bibfnamefont {S.}~\bibnamefont
  {Yuan}}, \bibinfo {author} {\bibfnamefont {M.~I.}\ \bibnamefont
  {Katsnelson}}, \bibinfo {author} {\bibfnamefont {H.-M.}\ \bibnamefont
  {Cheng}}, \bibinfo {author} {\bibfnamefont {W.}~\bibnamefont {Strupinski}},
  \bibinfo {author} {\bibfnamefont {L.~G.}\ \bibnamefont {Bulusheva}}, \bibinfo
  {author} {\bibfnamefont {A.~V.}\ \bibnamefont {Okotrub}}, \bibinfo {author}
  {\bibfnamefont {I.~V.}\ \bibnamefont {Grigorieva}}, \bibinfo {author}
  {\bibfnamefont {A.~N.}\ \bibnamefont {Grigorenko}}, \bibinfo {author}
  {\bibfnamefont {K.~S.}\ \bibnamefont {Novoselov}}, \ and\ \bibinfo {author}
  {\bibfnamefont {A.~K.}\ \bibnamefont {Geim}},\ }\href@noop {} {\bibfield
  {journal} {\bibinfo  {journal} {Small}\ }\textbf {\bibinfo {volume} {6}},\
  \bibinfo {pages} {2877} (\bibinfo {year} {2010})}\BibitemShut {NoStop}%
\bibitem [{\citenamefont {Cheng}\ \emph {et~al.}(2010)\citenamefont {Cheng},
  \citenamefont {Zou}, \citenamefont {Okino}, \citenamefont {Gutierrez},
  \citenamefont {Gupta}, \citenamefont {Shen}, \citenamefont {Eklund},
  \citenamefont {Sofo},\ and\ \citenamefont {Zhu}}]{PhysRevB.81.205435}%
  \BibitemOpen
  \bibfield  {author} {\bibinfo {author} {\bibfnamefont {S.-H.}\ \bibnamefont
  {Cheng}}, \bibinfo {author} {\bibfnamefont {K.}~\bibnamefont {Zou}}, \bibinfo
  {author} {\bibfnamefont {F.}~\bibnamefont {Okino}}, \bibinfo {author}
  {\bibfnamefont {H.~R.}\ \bibnamefont {Gutierrez}}, \bibinfo {author}
  {\bibfnamefont {A.}~\bibnamefont {Gupta}}, \bibinfo {author} {\bibfnamefont
  {N.}~\bibnamefont {Shen}}, \bibinfo {author} {\bibfnamefont {P.~C.}\
  \bibnamefont {Eklund}}, \bibinfo {author} {\bibfnamefont {J.~O.}\
  \bibnamefont {Sofo}}, \ and\ \bibinfo {author} {\bibfnamefont
  {J.}~\bibnamefont {Zhu}},\ }\href@noop {} {\bibfield  {journal} {\bibinfo
  {journal} {Phys. Rev. B}\ }\textbf {\bibinfo {volume} {81}},\ \bibinfo
  {pages} {205435} (\bibinfo {year} {2010})}\BibitemShut {NoStop}%
\bibitem [{\citenamefont {Robinson}\ \emph {et~al.}(2010)\citenamefont
  {Robinson}, \citenamefont {Burgess}, \citenamefont {Junkermeier},
  \citenamefont {Badescu}, \citenamefont {Reinecke}, \citenamefont {Perkins},
  \citenamefont {Zalalutdniov}, \citenamefont {Baldwin}, \citenamefont
  {Culbertson}, \citenamefont {Sheehan},\ and\ \citenamefont
  {Snow}}]{doi:10.1021/nl101437p}%
  \BibitemOpen
  \bibfield  {author} {\bibinfo {author} {\bibfnamefont {J.~T.}\ \bibnamefont
  {Robinson}}, \bibinfo {author} {\bibfnamefont {J.~S.}\ \bibnamefont
  {Burgess}}, \bibinfo {author} {\bibfnamefont {C.~E.}\ \bibnamefont
  {Junkermeier}}, \bibinfo {author} {\bibfnamefont {S.~C.}\ \bibnamefont
  {Badescu}}, \bibinfo {author} {\bibfnamefont {T.~L.}\ \bibnamefont
  {Reinecke}}, \bibinfo {author} {\bibfnamefont {F.~K.}\ \bibnamefont
  {Perkins}}, \bibinfo {author} {\bibfnamefont {M.~K.}\ \bibnamefont
  {Zalalutdniov}}, \bibinfo {author} {\bibfnamefont {J.~W.}\ \bibnamefont
  {Baldwin}}, \bibinfo {author} {\bibfnamefont {J.~C.}\ \bibnamefont
  {Culbertson}}, \bibinfo {author} {\bibfnamefont {P.~E.}\ \bibnamefont
  {Sheehan}}, \ and\ \bibinfo {author} {\bibfnamefont {E.~S.}\ \bibnamefont
  {Snow}},\ }\href@noop {} {\bibfield  {journal} {\bibinfo  {journal} {Nano
  Letters}\ }\textbf {\bibinfo {volume} {10}},\ \bibinfo {pages} {3001}
  (\bibinfo {year} {2010})}\BibitemShut {NoStop}%
\bibitem [{\citenamefont {Denis}(2013)}]{DENIS2013203}%
  \BibitemOpen
  \bibfield  {author} {\bibinfo {author} {\bibfnamefont {P.~A.}\ \bibnamefont
  {Denis}},\ }\href@noop {} {\bibfield  {journal} {\bibinfo  {journal}
  {Computational Materials Science}\ }\textbf {\bibinfo {volume} {67}},\
  \bibinfo {pages} {203 } (\bibinfo {year} {2013})}\BibitemShut {NoStop}%
\bibitem [{\citenamefont {Denis}\ \emph {et~al.}(2017)\citenamefont {Denis},
  \citenamefont {Huelmo},\ and\ \citenamefont {Iribarne}}]{DENIS201720}%
  \BibitemOpen
  \bibfield  {author} {\bibinfo {author} {\bibfnamefont {P.~A.}\ \bibnamefont
  {Denis}}, \bibinfo {author} {\bibfnamefont {C.~P.}\ \bibnamefont {Huelmo}}, \
  and\ \bibinfo {author} {\bibfnamefont {F.}~\bibnamefont {Iribarne}},\
  }\href@noop {} {\bibfield  {journal} {\bibinfo  {journal} {Computational
  Materials Science}\ }\textbf {\bibinfo {volume} {137}},\ \bibinfo {pages} {20
  } (\bibinfo {year} {2017})}\BibitemShut {NoStop}%
\bibitem [{\citenamefont {Son}\ \emph {et~al.}(2006{\natexlab{a}})\citenamefont
  {Son}, \citenamefont {Cohen},\ and\ \citenamefont
  {Louie}}]{PhysRevLett.97.216803}%
  \BibitemOpen
  \bibfield  {author} {\bibinfo {author} {\bibfnamefont {Y.-W.}\ \bibnamefont
  {Son}}, \bibinfo {author} {\bibfnamefont {M.~L.}\ \bibnamefont {Cohen}}, \
  and\ \bibinfo {author} {\bibfnamefont {S.~G.}\ \bibnamefont {Louie}},\
  }\href@noop {} {\bibfield  {journal} {\bibinfo  {journal} {Phys. Rev. Lett.}\
  }\textbf {\bibinfo {volume} {97}},\ \bibinfo {pages} {216803} (\bibinfo
  {year} {2006}{\natexlab{a}})}\BibitemShut {NoStop}%
\bibitem [{\citenamefont {Son}\ \emph {et~al.}(2006{\natexlab{b}})\citenamefont
  {Son}, \citenamefont {Cohen},\ and\ \citenamefont {Louie}}]{Son2006}%
  \BibitemOpen
  \bibfield  {author} {\bibinfo {author} {\bibfnamefont {Y.-W.}\ \bibnamefont
  {Son}}, \bibinfo {author} {\bibfnamefont {M.~L.}\ \bibnamefont {Cohen}}, \
  and\ \bibinfo {author} {\bibfnamefont {S.~G.}\ \bibnamefont {Louie}},\
  }\href@noop {} {\bibfield  {journal} {\bibinfo  {journal} {Nature}\ }\textbf
  {\bibinfo {volume} {444}},\ \bibinfo {pages} {347} (\bibinfo {year}
  {2006}{\natexlab{b}})}\BibitemShut {NoStop}%
\bibitem [{\citenamefont {Barone}\ \emph {et~al.}(2006)\citenamefont {Barone},
  \citenamefont {Hod},\ and\ \citenamefont {Scuseria}}]{doi:10.1021/nl0617033}%
  \BibitemOpen
  \bibfield  {author} {\bibinfo {author} {\bibfnamefont {V.}~\bibnamefont
  {Barone}}, \bibinfo {author} {\bibfnamefont {O.}~\bibnamefont {Hod}}, \ and\
  \bibinfo {author} {\bibfnamefont {G.~E.}\ \bibnamefont {Scuseria}},\
  }\href@noop {} {\bibfield  {journal} {\bibinfo  {journal} {Nano Letters}\
  }\textbf {\bibinfo {volume} {6}},\ \bibinfo {pages} {2748} (\bibinfo {year}
  {2006})}\BibitemShut {NoStop}%
\bibitem [{\citenamefont {Liu}\ \emph {et~al.}(2013)\citenamefont {Liu},
  \citenamefont {Ma}, \citenamefont {Shi}, \citenamefont {Zhou}, \citenamefont
  {Gong}, \citenamefont {Lei}, \citenamefont {Yang}, \citenamefont {Zhang},
  \citenamefont {Yu}, \citenamefont {Hackenberg}, \citenamefont {Babakhani},
  \citenamefont {Idrobo}, \citenamefont {Vajtai}, \citenamefont {Lou},\ and\
  \citenamefont {Ajayan}}]{Liu2013}%
  \BibitemOpen
  \bibfield  {author} {\bibinfo {author} {\bibfnamefont {Z.}~\bibnamefont
  {Liu}}, \bibinfo {author} {\bibfnamefont {L.}~\bibnamefont {Ma}}, \bibinfo
  {author} {\bibfnamefont {G.}~\bibnamefont {Shi}}, \bibinfo {author}
  {\bibfnamefont {W.}~\bibnamefont {Zhou}}, \bibinfo {author} {\bibfnamefont
  {Y.}~\bibnamefont {Gong}}, \bibinfo {author} {\bibfnamefont {S.}~\bibnamefont
  {Lei}}, \bibinfo {author} {\bibfnamefont {X.}~\bibnamefont {Yang}}, \bibinfo
  {author} {\bibfnamefont {J.}~\bibnamefont {Zhang}}, \bibinfo {author}
  {\bibfnamefont {J.}~\bibnamefont {Yu}}, \bibinfo {author} {\bibfnamefont
  {K.~P.}\ \bibnamefont {Hackenberg}}, \bibinfo {author} {\bibfnamefont
  {A.}~\bibnamefont {Babakhani}}, \bibinfo {author} {\bibfnamefont {J.-C.}\
  \bibnamefont {Idrobo}}, \bibinfo {author} {\bibfnamefont {R.}~\bibnamefont
  {Vajtai}}, \bibinfo {author} {\bibfnamefont {J.}~\bibnamefont {Lou}}, \ and\
  \bibinfo {author} {\bibfnamefont {P.~M.}\ \bibnamefont {Ajayan}},\
  }\href@noop {} {\bibfield  {journal} {\bibinfo  {journal} {Nature
  Nanotechnology}\ }\textbf {\bibinfo {volume} {8}},\ \bibinfo {pages} {119}
  (\bibinfo {year} {2013})}\BibitemShut {NoStop}%
\bibitem [{\citenamefont {Nguyen}\ \emph {et~al.}(2011)\citenamefont {Nguyen},
  \citenamefont {Otani},\ and\ \citenamefont {Okada}}]{PhysRevLett.106.106801}%
  \BibitemOpen
  \bibfield  {author} {\bibinfo {author} {\bibfnamefont {T.~C.}\ \bibnamefont
  {Nguyen}}, \bibinfo {author} {\bibfnamefont {M.}~\bibnamefont {Otani}}, \
  and\ \bibinfo {author} {\bibfnamefont {S.}~\bibnamefont {Okada}},\
  }\href@noop {} {\bibfield  {journal} {\bibinfo  {journal} {Phys. Rev. Lett.}\
  }\textbf {\bibinfo {volume} {106}},\ \bibinfo {pages} {106801} (\bibinfo
  {year} {2011})}\BibitemShut {NoStop}%
\bibitem [{\citenamefont {Zhou}\ \emph {et~al.}(2007)\citenamefont {Zhou},
  \citenamefont {Gweon}, \citenamefont {Fedorov}, \citenamefont {First},
  \citenamefont {de~Heer}, \citenamefont {Lee}, \citenamefont {Guinea},
  \citenamefont {Castro~Neto},\ and\ \citenamefont {Lanzara}}]{Zhou2007}%
  \BibitemOpen
  \bibfield  {author} {\bibinfo {author} {\bibfnamefont {S.~Y.}\ \bibnamefont
  {Zhou}}, \bibinfo {author} {\bibfnamefont {G.-H.}\ \bibnamefont {Gweon}},
  \bibinfo {author} {\bibfnamefont {A.~V.}\ \bibnamefont {Fedorov}}, \bibinfo
  {author} {\bibfnamefont {P.~N.}\ \bibnamefont {First}}, \bibinfo {author}
  {\bibfnamefont {W.~A.}\ \bibnamefont {de~Heer}}, \bibinfo {author}
  {\bibfnamefont {D.-H.}\ \bibnamefont {Lee}}, \bibinfo {author} {\bibfnamefont
  {F.}~\bibnamefont {Guinea}}, \bibinfo {author} {\bibfnamefont {A.~H.}\
  \bibnamefont {Castro~Neto}}, \ and\ \bibinfo {author} {\bibfnamefont
  {A.}~\bibnamefont {Lanzara}},\ }\href@noop {} {\bibfield  {journal} {\bibinfo
   {journal} {Nature Materials}\ }\textbf {\bibinfo {volume} {6}},\ \bibinfo
  {pages} {770} (\bibinfo {year} {2007})}\BibitemShut {NoStop}%
\bibitem [{\citenamefont {Giovannetti}\ \emph {et~al.}(2007)\citenamefont
  {Giovannetti}, \citenamefont {Khomyakov}, \citenamefont {Brocks},
  \citenamefont {Kelly},\ and\ \citenamefont {van~den
  Brink}}]{PhysRevB.76.073103}%
  \BibitemOpen
  \bibfield  {author} {\bibinfo {author} {\bibfnamefont {G.}~\bibnamefont
  {Giovannetti}}, \bibinfo {author} {\bibfnamefont {P.~A.}\ \bibnamefont
  {Khomyakov}}, \bibinfo {author} {\bibfnamefont {G.}~\bibnamefont {Brocks}},
  \bibinfo {author} {\bibfnamefont {P.~J.}\ \bibnamefont {Kelly}}, \ and\
  \bibinfo {author} {\bibfnamefont {J.}~\bibnamefont {van~den Brink}},\
  }\href@noop {} {\bibfield  {journal} {\bibinfo  {journal} {Phys. Rev. B}\
  }\textbf {\bibinfo {volume} {76}},\ \bibinfo {pages} {073103} (\bibinfo
  {year} {2007})}\BibitemShut {NoStop}%
\bibitem [{\citenamefont {Skomski}\ \emph {et~al.}(2014)\citenamefont
  {Skomski}, \citenamefont {Dowben}, \citenamefont {Sky~Driver},\ and\
  \citenamefont {Kelber}}]{C4MH00124A}%
  \BibitemOpen
  \bibfield  {author} {\bibinfo {author} {\bibfnamefont {R.}~\bibnamefont
  {Skomski}}, \bibinfo {author} {\bibfnamefont {P.~A.}\ \bibnamefont {Dowben}},
  \bibinfo {author} {\bibfnamefont {M.}~\bibnamefont {Sky~Driver}}, \ and\
  \bibinfo {author} {\bibfnamefont {J.~A.}\ \bibnamefont {Kelber}},\
  }\href@noop {} {\bibfield  {journal} {\bibinfo  {journal} {Mater. Horiz.}\
  }\textbf {\bibinfo {volume} {1}},\ \bibinfo {pages} {563} (\bibinfo {year}
  {2014})}\BibitemShut {NoStop}%
\bibitem [{\citenamefont {Denis}(2010)}]{DENIS2010251}%
  \BibitemOpen
  \bibfield  {author} {\bibinfo {author} {\bibfnamefont {P.~A.}\ \bibnamefont
  {Denis}},\ }\href@noop {} {\bibfield  {journal} {\bibinfo  {journal}
  {Chemical Physics Letters}\ }\textbf {\bibinfo {volume} {492}},\ \bibinfo
  {pages} {251 } (\bibinfo {year} {2010})}\BibitemShut {NoStop}%
\bibitem [{\citenamefont {Denis}(2014)}]{doi:10.1002/cphc.201402608}%
  \BibitemOpen
  \bibfield  {author} {\bibinfo {author} {\bibfnamefont {P.~A.}\ \bibnamefont
  {Denis}},\ }\href@noop {} {\bibfield  {journal} {\bibinfo  {journal}
  {ChemPhysChem}\ }\textbf {\bibinfo {volume} {15}},\ \bibinfo {pages} {3994}
  (\bibinfo {year} {2014})}\BibitemShut {NoStop}%
\bibitem [{\citenamefont {Denis}\ and\ \citenamefont
  {Iribarne}(2016{\natexlab{a}})}]{DENIS2016152}%
  \BibitemOpen
  \bibfield  {author} {\bibinfo {author} {\bibfnamefont {P.~A.}\ \bibnamefont
  {Denis}}\ and\ \bibinfo {author} {\bibfnamefont {F.}~\bibnamefont
  {Iribarne}},\ }\href@noop {} {\bibfield  {journal} {\bibinfo  {journal}
  {Chemical Physics Letters}\ }\textbf {\bibinfo {volume} {658}},\ \bibinfo
  {pages} {152 } (\bibinfo {year} {2016}{\natexlab{a}})}\BibitemShut {NoStop}%
\bibitem [{\citenamefont {Menezes}\ \emph {et~al.}(2010)\citenamefont
  {Menezes}, \citenamefont {Capaz},\ and\ \citenamefont
  {Faria}}]{PhysRevB.82.245414}%
  \BibitemOpen
  \bibfield  {author} {\bibinfo {author} {\bibfnamefont {M.~G.}\ \bibnamefont
  {Menezes}}, \bibinfo {author} {\bibfnamefont {R.~B.}\ \bibnamefont {Capaz}},
  \ and\ \bibinfo {author} {\bibfnamefont {J.~L.~B.}\ \bibnamefont {Faria}},\
  }\href@noop {} {\bibfield  {journal} {\bibinfo  {journal} {Phys. Rev. B}\
  }\textbf {\bibinfo {volume} {82}},\ \bibinfo {pages} {245414} (\bibinfo
  {year} {2010})}\BibitemShut {NoStop}%
\bibitem [{\citenamefont {Fujimoto}(2015)}]{Fujimoto2015}%
  \BibitemOpen
  \bibfield  {author} {\bibinfo {author} {\bibfnamefont {Y.}~\bibnamefont
  {Fujimoto}},\ }\href@noop {} {\bibfield  {journal} {\bibinfo  {journal}
  {Advances in Condensed Matter Physics}\ }\textbf {\bibinfo {volume} {2015}},\
  \bibinfo {pages} {571490} (\bibinfo {year} {2015})}\BibitemShut {NoStop}%
\bibitem [{\citenamefont {Nemnes}\ \emph {et~al.}(2018)\citenamefont {Nemnes},
  \citenamefont {Mitran}, \citenamefont {Manolescu},\ and\ \citenamefont
  {Dragoman}}]{NEMNES2018175}%
  \BibitemOpen
  \bibfield  {author} {\bibinfo {author} {\bibfnamefont {G.}~\bibnamefont
  {Nemnes}}, \bibinfo {author} {\bibfnamefont {T.}~\bibnamefont {Mitran}},
  \bibinfo {author} {\bibfnamefont {A.}~\bibnamefont {Manolescu}}, \ and\
  \bibinfo {author} {\bibfnamefont {D.}~\bibnamefont {Dragoman}},\ }\href@noop
  {} {\bibfield  {journal} {\bibinfo  {journal} {Computational Materials
  Science}\ }\textbf {\bibinfo {volume} {155}},\ \bibinfo {pages} {175 }
  (\bibinfo {year} {2018})}\BibitemShut {NoStop}%
\bibitem [{\citenamefont {Boukhvalov}\ and\ \citenamefont
  {Katsnelson}(2008)}]{PhysRevB.78.085413}%
  \BibitemOpen
  \bibfield  {author} {\bibinfo {author} {\bibfnamefont {D.~W.}\ \bibnamefont
  {Boukhvalov}}\ and\ \bibinfo {author} {\bibfnamefont {M.~I.}\ \bibnamefont
  {Katsnelson}},\ }\href@noop {} {\bibfield  {journal} {\bibinfo  {journal}
  {Phys. Rev. B}\ }\textbf {\bibinfo {volume} {78}},\ \bibinfo {pages} {085413}
  (\bibinfo {year} {2008})}\BibitemShut {NoStop}%
\bibitem [{\citenamefont {Hu}\ and\ \citenamefont {Gerber}(2014)}]{HU201475}%
  \BibitemOpen
  \bibfield  {author} {\bibinfo {author} {\bibfnamefont {T.}~\bibnamefont
  {Hu}}\ and\ \bibinfo {author} {\bibfnamefont {I.~C.}\ \bibnamefont
  {Gerber}},\ }\href@noop {} {\bibfield  {journal} {\bibinfo  {journal}
  {Chemical Physics Letters}\ }\textbf {\bibinfo {volume} {616-617}},\ \bibinfo
  {pages} {75 } (\bibinfo {year} {2014})}\BibitemShut {NoStop}%
\bibitem [{\citenamefont {Ohta}\ \emph {et~al.}(2006)\citenamefont {Ohta},
  \citenamefont {Bostwick}, \citenamefont {Seyller}, \citenamefont {Horn},\
  and\ \citenamefont {Rotenberg}}]{Ohta951}%
  \BibitemOpen
  \bibfield  {author} {\bibinfo {author} {\bibfnamefont {T.}~\bibnamefont
  {Ohta}}, \bibinfo {author} {\bibfnamefont {A.}~\bibnamefont {Bostwick}},
  \bibinfo {author} {\bibfnamefont {T.}~\bibnamefont {Seyller}}, \bibinfo
  {author} {\bibfnamefont {K.}~\bibnamefont {Horn}}, \ and\ \bibinfo {author}
  {\bibfnamefont {E.}~\bibnamefont {Rotenberg}},\ }\href@noop {} {\bibfield
  {journal} {\bibinfo  {journal} {Science}\ }\textbf {\bibinfo {volume}
  {313}},\ \bibinfo {pages} {951} (\bibinfo {year} {2006})}\BibitemShut
  {NoStop}%
\bibitem [{\citenamefont {Tang}\ \emph {et~al.}(2017)\citenamefont {Tang},
  \citenamefont {Wu}, \citenamefont {Xie}, \citenamefont {Li},\ and\
  \citenamefont {Gu}}]{C7RA01134B}%
  \BibitemOpen
  \bibfield  {author} {\bibinfo {author} {\bibfnamefont {S.}~\bibnamefont
  {Tang}}, \bibinfo {author} {\bibfnamefont {W.}~\bibnamefont {Wu}}, \bibinfo
  {author} {\bibfnamefont {X.}~\bibnamefont {Xie}}, \bibinfo {author}
  {\bibfnamefont {X.}~\bibnamefont {Li}}, \ and\ \bibinfo {author}
  {\bibfnamefont {J.}~\bibnamefont {Gu}},\ }\href@noop {} {\bibfield  {journal}
  {\bibinfo  {journal} {RSC Adv.}\ }\textbf {\bibinfo {volume} {7}},\ \bibinfo
  {pages} {9862} (\bibinfo {year} {2017})}\BibitemShut {NoStop}%
\bibitem [{\citenamefont {Mao}\ \emph {et~al.}(2010)\citenamefont {Mao},
  \citenamefont {Stocks},\ and\ \citenamefont {Zhong}}]{1367-2630-12-3-033046}%
  \BibitemOpen
  \bibfield  {author} {\bibinfo {author} {\bibfnamefont {Y.}~\bibnamefont
  {Mao}}, \bibinfo {author} {\bibfnamefont {G.~M.}\ \bibnamefont {Stocks}}, \
  and\ \bibinfo {author} {\bibfnamefont {J.}~\bibnamefont {Zhong}},\
  }\href@noop {} {\bibfield  {journal} {\bibinfo  {journal} {New Journal of
  Physics}\ }\textbf {\bibinfo {volume} {12}},\ \bibinfo {pages} {033046}
  (\bibinfo {year} {2010})}\BibitemShut {NoStop}%
\bibitem [{\citenamefont {Mao}\ and\ \citenamefont
  {Zhong}(2008)}]{0957-4484-19-20-205708}%
  \BibitemOpen
  \bibfield  {author} {\bibinfo {author} {\bibfnamefont {Y.}~\bibnamefont
  {Mao}}\ and\ \bibinfo {author} {\bibfnamefont {J.}~\bibnamefont {Zhong}},\
  }\href@noop {} {\bibfield  {journal} {\bibinfo  {journal} {Nanotechnology}\
  }\textbf {\bibinfo {volume} {19}},\ \bibinfo {pages} {205708} (\bibinfo
  {year} {2008})}\BibitemShut {NoStop}%
\bibitem [{\citenamefont {Nemnes}(2012)}]{NEMNES1}%
  \BibitemOpen
  \bibfield  {author} {\bibinfo {author} {\bibfnamefont {G.~A.}\ \bibnamefont
  {Nemnes}},\ }\href@noop {} {\bibfield  {journal} {\bibinfo  {journal}
  {Journal of Nanomaterials}\ }\textbf {\bibinfo {volume} {2012}},\ \bibinfo
  {pages} {748639} (\bibinfo {year} {2012})}\BibitemShut {NoStop}%
\bibitem [{\citenamefont {Nemnes}\ and\ \citenamefont
  {Antohe}(2013)}]{NEMNES20131347}%
  \BibitemOpen
  \bibfield  {author} {\bibinfo {author} {\bibfnamefont {G.~A.}\ \bibnamefont
  {Nemnes}}\ and\ \bibinfo {author} {\bibfnamefont {S.}~\bibnamefont
  {Antohe}},\ }\href@noop {} {\bibfield  {journal} {\bibinfo  {journal}
  {Materials Science and Engineering: B}\ }\textbf {\bibinfo {volume} {178}},\
  \bibinfo {pages} {1347 } (\bibinfo {year} {2013})}\BibitemShut {NoStop}%
\bibitem [{\citenamefont {Park}\ \emph
  {et~al.}(2015{\natexlab{a}})\citenamefont {Park}, \citenamefont {Zhao},
  \citenamefont {Shon}, \citenamefont {Yoon}, \citenamefont {Lee},
  \citenamefont {Song}, \citenamefont {Lee},\ and\ \citenamefont
  {Kim}}]{C5TC00051C}%
  \BibitemOpen
  \bibfield  {author} {\bibinfo {author} {\bibfnamefont {C.-S.}\ \bibnamefont
  {Park}}, \bibinfo {author} {\bibfnamefont {Y.}~\bibnamefont {Zhao}}, \bibinfo
  {author} {\bibfnamefont {Y.}~\bibnamefont {Shon}}, \bibinfo {author}
  {\bibfnamefont {I.~T.}\ \bibnamefont {Yoon}}, \bibinfo {author}
  {\bibfnamefont {C.~J.}\ \bibnamefont {Lee}}, \bibinfo {author} {\bibfnamefont
  {J.~D.}\ \bibnamefont {Song}}, \bibinfo {author} {\bibfnamefont
  {H.}~\bibnamefont {Lee}}, \ and\ \bibinfo {author} {\bibfnamefont {E.~K.}\
  \bibnamefont {Kim}},\ }\href@noop {} {\bibfield  {journal} {\bibinfo
  {journal} {J. Mater. Chem. C}\ }\textbf {\bibinfo {volume} {3}},\ \bibinfo
  {pages} {4235} (\bibinfo {year} {2015}{\natexlab{a}})}\BibitemShut {NoStop}%
\bibitem [{\citenamefont {Samuels}\ and\ \citenamefont
  {Carey}(2013)}]{doi:10.1021/nn400340q}%
  \BibitemOpen
  \bibfield  {author} {\bibinfo {author} {\bibfnamefont {A.~J.}\ \bibnamefont
  {Samuels}}\ and\ \bibinfo {author} {\bibfnamefont {J.~D.}\ \bibnamefont
  {Carey}},\ }\href@noop {} {\bibfield  {journal} {\bibinfo  {journal} {ACS
  Nano}\ }\textbf {\bibinfo {volume} {7}},\ \bibinfo {pages} {2790} (\bibinfo
  {year} {2013})}\BibitemShut {NoStop}%
\bibitem [{\citenamefont {Uchiyama}\ \emph {et~al.}(2017)\citenamefont
  {Uchiyama}, \citenamefont {Goto}, \citenamefont {Akiyoshi}, \citenamefont
  {Eguchi}, \citenamefont {Nishikawa}, \citenamefont {Osada},\ and\
  \citenamefont {Kubozono}}]{Uchiyama2017}%
  \BibitemOpen
  \bibfield  {author} {\bibinfo {author} {\bibfnamefont {T.}~\bibnamefont
  {Uchiyama}}, \bibinfo {author} {\bibfnamefont {H.}~\bibnamefont {Goto}},
  \bibinfo {author} {\bibfnamefont {H.}~\bibnamefont {Akiyoshi}}, \bibinfo
  {author} {\bibfnamefont {R.}~\bibnamefont {Eguchi}}, \bibinfo {author}
  {\bibfnamefont {T.}~\bibnamefont {Nishikawa}}, \bibinfo {author}
  {\bibfnamefont {H.}~\bibnamefont {Osada}}, \ and\ \bibinfo {author}
  {\bibfnamefont {Y.}~\bibnamefont {Kubozono}},\ }\href@noop {} {\bibfield
  {journal} {\bibinfo  {journal} {Scientific Reports}\ }\textbf {\bibinfo
  {volume} {7}},\ \bibinfo {pages} {11322} (\bibinfo {year}
  {2017})}\BibitemShut {NoStop}%
\bibitem [{\citenamefont {Yamada}\ \emph {et~al.}(2018)\citenamefont {Yamada},
  \citenamefont {Okigawa},\ and\ \citenamefont
  {Hasegawa}}]{doi:10.1063/1.5012808}%
  \BibitemOpen
  \bibfield  {author} {\bibinfo {author} {\bibfnamefont {T.}~\bibnamefont
  {Yamada}}, \bibinfo {author} {\bibfnamefont {Y.}~\bibnamefont {Okigawa}}, \
  and\ \bibinfo {author} {\bibfnamefont {M.}~\bibnamefont {Hasegawa}},\
  }\href@noop {} {\bibfield  {journal} {\bibinfo  {journal} {Applied Physics
  Letters}\ }\textbf {\bibinfo {volume} {112}},\ \bibinfo {pages} {043105}
  (\bibinfo {year} {2018})}\BibitemShut {NoStop}%
\bibitem [{\citenamefont {Soler}\ \emph {et~al.}(2002)\citenamefont {Soler},
  \citenamefont {Artacho}, \citenamefont {Gale}, \citenamefont {Garcia},
  \citenamefont {Junquera}, \citenamefont {Ordejon},\ and\ \citenamefont
  {Sanchez-Portal}}]{0953-8984-14-11-302}%
  \BibitemOpen
  \bibfield  {author} {\bibinfo {author} {\bibfnamefont {J.~M.}\ \bibnamefont
  {Soler}}, \bibinfo {author} {\bibfnamefont {E.}~\bibnamefont {Artacho}},
  \bibinfo {author} {\bibfnamefont {J.~D.}\ \bibnamefont {Gale}}, \bibinfo
  {author} {\bibfnamefont {A.}~\bibnamefont {Garcia}}, \bibinfo {author}
  {\bibfnamefont {J.}~\bibnamefont {Junquera}}, \bibinfo {author}
  {\bibfnamefont {P.}~\bibnamefont {Ordejon}}, \ and\ \bibinfo {author}
  {\bibfnamefont {D.}~\bibnamefont {Sanchez-Portal}},\ }\href@noop {}
  {\bibfield  {journal} {\bibinfo  {journal} {Journal of Physics: Condensed
  Matter}\ }\textbf {\bibinfo {volume} {14}},\ \bibinfo {pages} {2745}
  (\bibinfo {year} {2002})}\BibitemShut {NoStop}%
\bibitem [{\citenamefont {Dion}\ \emph {et~al.}(2004)\citenamefont {Dion},
  \citenamefont {Rydberg}, \citenamefont {Schr\"oder}, \citenamefont
  {Langreth},\ and\ \citenamefont {Lundqvist}}]{PhysRevLett.92.246401}%
  \BibitemOpen
  \bibfield  {author} {\bibinfo {author} {\bibfnamefont {M.}~\bibnamefont
  {Dion}}, \bibinfo {author} {\bibfnamefont {H.}~\bibnamefont {Rydberg}},
  \bibinfo {author} {\bibfnamefont {E.}~\bibnamefont {Schr\"oder}}, \bibinfo
  {author} {\bibfnamefont {D.~C.}\ \bibnamefont {Langreth}}, \ and\ \bibinfo
  {author} {\bibfnamefont {B.~I.}\ \bibnamefont {Lundqvist}},\ }\href@noop {}
  {\bibfield  {journal} {\bibinfo  {journal} {Phys. Rev. Lett.}\ }\textbf
  {\bibinfo {volume} {92}},\ \bibinfo {pages} {246401} (\bibinfo {year}
  {2004})}\BibitemShut {NoStop}%
\bibitem [{\citenamefont {Birowska}\ \emph {et~al.}(2011)\citenamefont
  {Birowska}, \citenamefont {Milowska},\ and\ \citenamefont
  {Majewski}}]{Birowska_Milowska_Majewski_2011}%
  \BibitemOpen
  \bibfield  {author} {\bibinfo {author} {\bibfnamefont {M.}~\bibnamefont
  {Birowska}}, \bibinfo {author} {\bibfnamefont {K.}~\bibnamefont {Milowska}},
  \ and\ \bibinfo {author} {\bibfnamefont {J.}~\bibnamefont {Majewski}},\
  }\href@noop {} {\bibfield  {journal} {\bibinfo  {journal} {Acta Physica
  Polonica A}\ }\textbf {\bibinfo {volume} {120}},\ \bibinfo {pages}
  {845–848} (\bibinfo {year} {2011})}\BibitemShut {NoStop}%
\bibitem [{\citenamefont {Troullier}\ and\ \citenamefont
  {Martins}(1991)}]{PhysRevB.43.1993}%
  \BibitemOpen
  \bibfield  {author} {\bibinfo {author} {\bibfnamefont {N.}~\bibnamefont
  {Troullier}}\ and\ \bibinfo {author} {\bibfnamefont {J.~L.}\ \bibnamefont
  {Martins}},\ }\href@noop {} {\bibfield  {journal} {\bibinfo  {journal} {Phys.
  Rev. B}\ }\textbf {\bibinfo {volume} {43}},\ \bibinfo {pages} {1993}
  (\bibinfo {year} {1991})}\BibitemShut {NoStop}%
\bibitem [{\citenamefont {Park}\ \emph
  {et~al.}(2015{\natexlab{b}})\citenamefont {Park}, \citenamefont {Ryou},
  \citenamefont {Hong}, \citenamefont {Sumpter}, \citenamefont {Kim},\ and\
  \citenamefont {Yoon}}]{PhysRevLett.115.015502}%
  \BibitemOpen
  \bibfield  {author} {\bibinfo {author} {\bibfnamefont {C.}~\bibnamefont
  {Park}}, \bibinfo {author} {\bibfnamefont {J.}~\bibnamefont {Ryou}}, \bibinfo
  {author} {\bibfnamefont {S.}~\bibnamefont {Hong}}, \bibinfo {author}
  {\bibfnamefont {B.~G.}\ \bibnamefont {Sumpter}}, \bibinfo {author}
  {\bibfnamefont {G.}~\bibnamefont {Kim}}, \ and\ \bibinfo {author}
  {\bibfnamefont {M.}~\bibnamefont {Yoon}},\ }\href@noop {} {\bibfield
  {journal} {\bibinfo  {journal} {Phys. Rev. Lett.}\ }\textbf {\bibinfo
  {volume} {115}},\ \bibinfo {pages} {015502} (\bibinfo {year}
  {2015}{\natexlab{b}})}\BibitemShut {NoStop}%
\bibitem [{\citenamefont {Denis}\ and\ \citenamefont
  {Iribarne}(2016{\natexlab{b}})}]{C6CP02481E}%
  \BibitemOpen
  \bibfield  {author} {\bibinfo {author} {\bibfnamefont {P.~A.}\ \bibnamefont
  {Denis}}\ and\ \bibinfo {author} {\bibfnamefont {F.}~\bibnamefont
  {Iribarne}},\ }\href@noop {} {\bibfield  {journal} {\bibinfo  {journal}
  {Phys. Chem. Chem. Phys.}\ }\textbf {\bibinfo {volume} {18}},\ \bibinfo
  {pages} {24693} (\bibinfo {year} {2016}{\natexlab{b}})}\BibitemShut {NoStop}%
\bibitem [{\citenamefont {Mousavi}\ \emph {et~al.}(2018)\citenamefont
  {Mousavi}, \citenamefont {Khodadadi},\ and\ \citenamefont
  {Grabowski}}]{MOUSAVI201890}%
  \BibitemOpen
  \bibfield  {author} {\bibinfo {author} {\bibfnamefont {H.}~\bibnamefont
  {Mousavi}}, \bibinfo {author} {\bibfnamefont {J.}~\bibnamefont {Khodadadi}},
  \ and\ \bibinfo {author} {\bibfnamefont {M.}~\bibnamefont {Grabowski}},\
  }\href@noop {} {\bibfield  {journal} {\bibinfo  {journal} {Physica B:
  Condensed Matter}\ }\textbf {\bibinfo {volume} {530}},\ \bibinfo {pages} {90}
  (\bibinfo {year} {2018})}\BibitemShut {NoStop}%
\bibitem [{\citenamefont {Cutler}\ and\ \citenamefont
  {Mott}(1969)}]{PhysRev.181.1336}%
  \BibitemOpen
  \bibfield  {author} {\bibinfo {author} {\bibfnamefont {M.}~\bibnamefont
  {Cutler}}\ and\ \bibinfo {author} {\bibfnamefont {N.~F.}\ \bibnamefont
  {Mott}},\ }\href@noop {} {\bibfield  {journal} {\bibinfo  {journal} {Phys.
  Rev.}\ }\textbf {\bibinfo {volume} {181}},\ \bibinfo {pages} {1336} (\bibinfo
  {year} {1969})}\BibitemShut {NoStop}%
\bibitem [{\citenamefont {Nemnes}\ \emph {et~al.}(2010)\citenamefont {Nemnes},
  \citenamefont {Ion},\ and\ \citenamefont {Antohe}}]{NEMNES20101613}%
  \BibitemOpen
  \bibfield  {author} {\bibinfo {author} {\bibfnamefont {G.}~\bibnamefont
  {Nemnes}}, \bibinfo {author} {\bibfnamefont {L.}~\bibnamefont {Ion}}, \ and\
  \bibinfo {author} {\bibfnamefont {S.}~\bibnamefont {Antohe}},\ }\href@noop {}
  {\bibfield  {journal} {\bibinfo  {journal} {Physica E: Low-dimensional
  Systems and Nanostructures}\ }\textbf {\bibinfo {volume} {42}},\ \bibinfo
  {pages} {1613 } (\bibinfo {year} {2010})}\BibitemShut {NoStop}%
\bibitem [{\citenamefont {Svensson}\ \emph {et~al.}(2012)\citenamefont
  {Svensson}, \citenamefont {Persson}, \citenamefont {Hoffmann}, \citenamefont
  {Nakpathomkun}, \citenamefont {Nilsson}, \citenamefont {Xu}, \citenamefont
  {Samuelson},\ and\ \citenamefont {Linke}}]{1367-2630-14-3-033041}%
  \BibitemOpen
  \bibfield  {author} {\bibinfo {author} {\bibfnamefont {S.~F.}\ \bibnamefont
  {Svensson}}, \bibinfo {author} {\bibfnamefont {A.~I.}\ \bibnamefont
  {Persson}}, \bibinfo {author} {\bibfnamefont {E.~A.}\ \bibnamefont
  {Hoffmann}}, \bibinfo {author} {\bibfnamefont {N.}~\bibnamefont
  {Nakpathomkun}}, \bibinfo {author} {\bibfnamefont {H.~A.}\ \bibnamefont
  {Nilsson}}, \bibinfo {author} {\bibfnamefont {H.~Q.}\ \bibnamefont {Xu}},
  \bibinfo {author} {\bibfnamefont {L.}~\bibnamefont {Samuelson}}, \ and\
  \bibinfo {author} {\bibfnamefont {H.}~\bibnamefont {Linke}},\ }\href@noop {}
  {\bibfield  {journal} {\bibinfo  {journal} {New Journal of Physics}\ }\textbf
  {\bibinfo {volume} {14}},\ \bibinfo {pages} {033041} (\bibinfo {year}
  {2012})}\BibitemShut {NoStop}%
\bibitem [{\citenamefont {Erlingsson}\ \emph {et~al.}(2017)\citenamefont
  {Erlingsson}, \citenamefont {Manolescu}, \citenamefont {Nemnes},
  \citenamefont {Bardarson},\ and\ \citenamefont
  {Sanchez}}]{PhysRevLett.119.036804}%
  \BibitemOpen
  \bibfield  {author} {\bibinfo {author} {\bibfnamefont {S.~I.}\ \bibnamefont
  {Erlingsson}}, \bibinfo {author} {\bibfnamefont {A.}~\bibnamefont
  {Manolescu}}, \bibinfo {author} {\bibfnamefont {G.~A.}\ \bibnamefont
  {Nemnes}}, \bibinfo {author} {\bibfnamefont {J.~H.}\ \bibnamefont
  {Bardarson}}, \ and\ \bibinfo {author} {\bibfnamefont {D.}~\bibnamefont
  {Sanchez}},\ }\href@noop {} {\bibfield  {journal} {\bibinfo  {journal} {Phys.
  Rev. Lett.}\ }\textbf {\bibinfo {volume} {119}},\ \bibinfo {pages} {036804}
  (\bibinfo {year} {2017})}\BibitemShut {NoStop}%
\bibitem [{\citenamefont {Xu}\ and\ \citenamefont
  {Verstraete}(2014)}]{PhysRevLett.112.196603}%
  \BibitemOpen
  \bibfield  {author} {\bibinfo {author} {\bibfnamefont {B.}~\bibnamefont
  {Xu}}\ and\ \bibinfo {author} {\bibfnamefont {M.~J.}\ \bibnamefont
  {Verstraete}},\ }\href@noop {} {\bibfield  {journal} {\bibinfo  {journal}
  {Phys. Rev. Lett.}\ }\textbf {\bibinfo {volume} {112}},\ \bibinfo {pages}
  {196603} (\bibinfo {year} {2014})}\BibitemShut {NoStop}%
\end{thebibliography}%

\clearpage
\onecolumngrid

\appendix*
\section{Supplementary Material}

\renewcommand{\thefigure}{S\arabic{figure}}
\setcounter{figure}{0}

\renewcommand{\thetable}{S\arabic{table}}
\setcounter{table}{0}

\vspace{2cm}

\begin{figure*}[h]
\centering
\includegraphics[width=4.5cm]{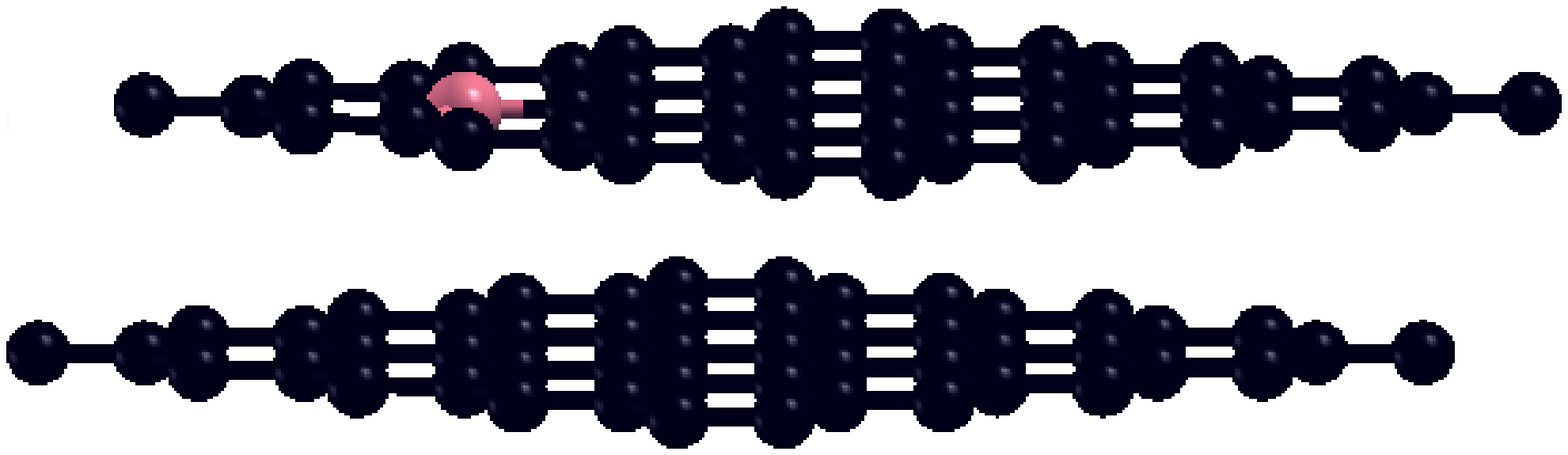} \hspace*{0.5cm}
\includegraphics[width=4.5cm]{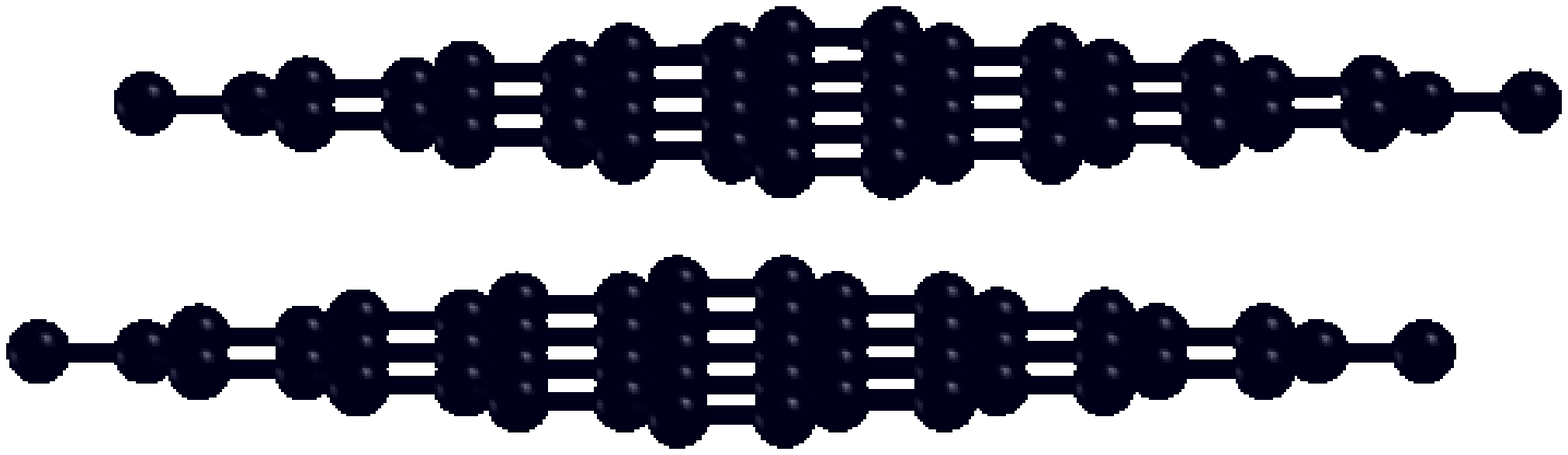} \hspace*{0.5cm}
\includegraphics[width=4.5cm]{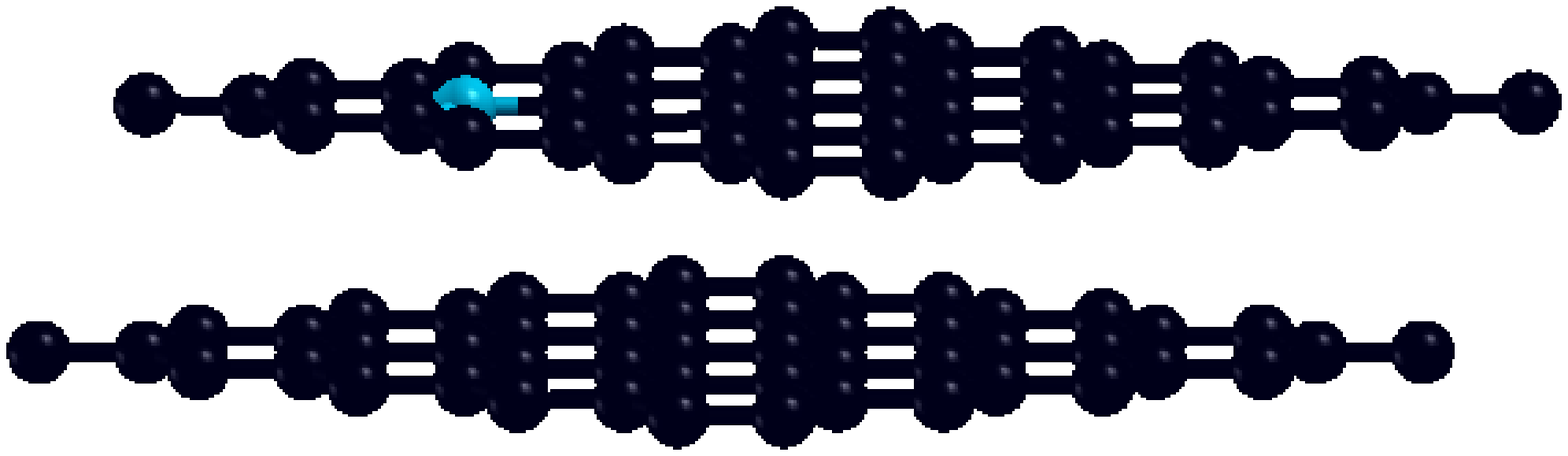} \vspace*{1cm}\\
\includegraphics[width=4.5cm]{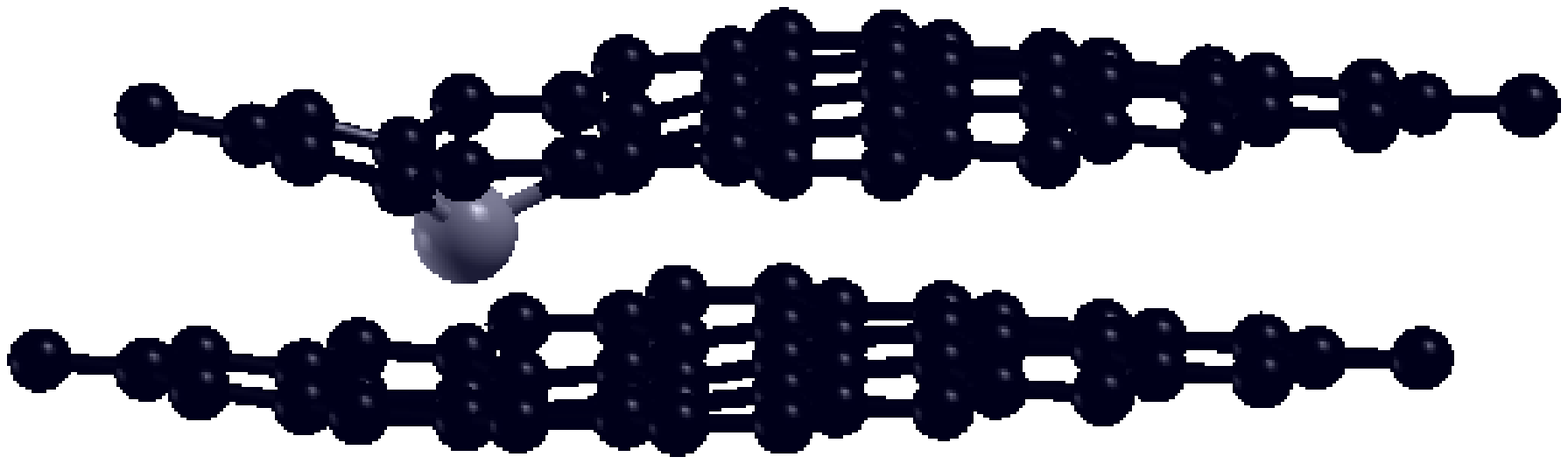} \hspace*{0.5cm}
\includegraphics[width=4.5cm]{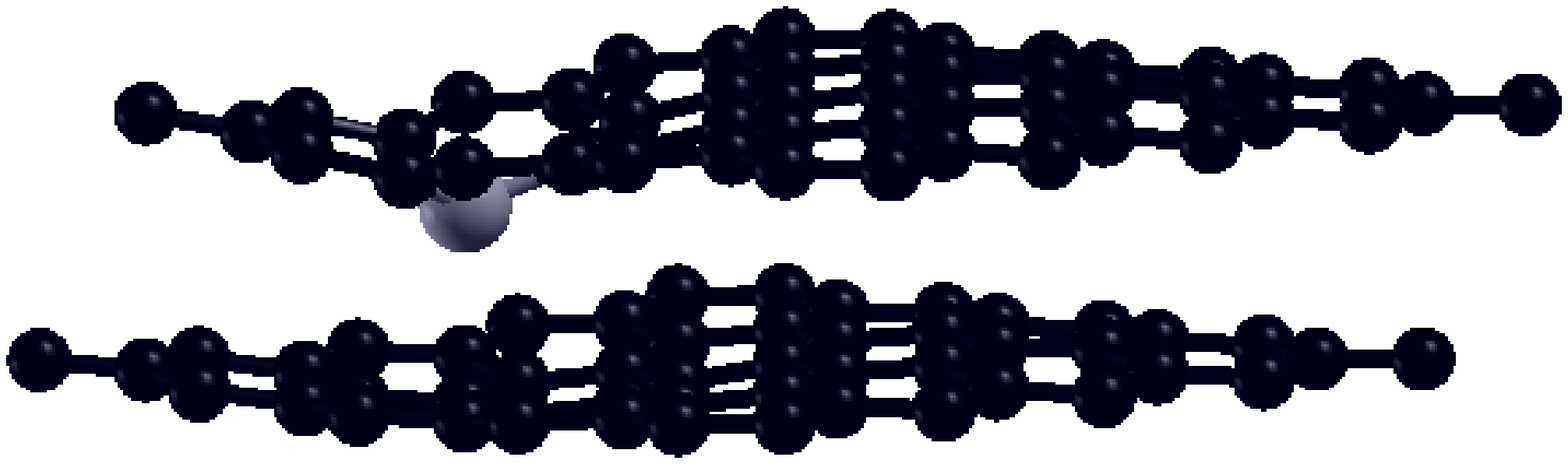} \hspace*{0.5cm}
\includegraphics[width=4.5cm]{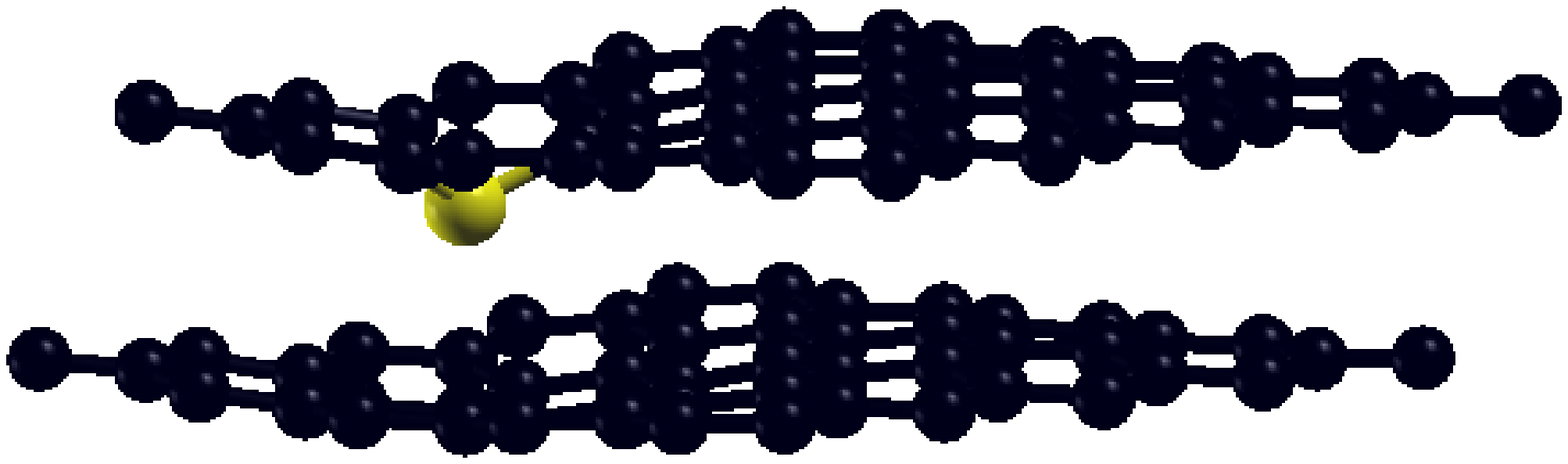} \vspace*{1cm}\\
\includegraphics[width=4.5cm]{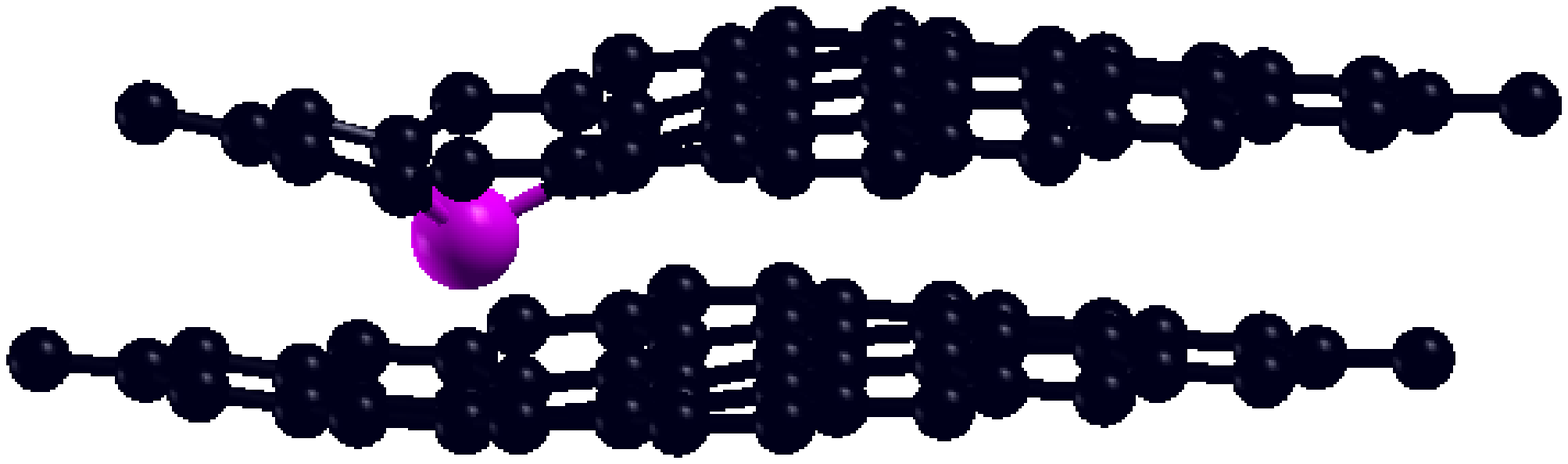} \hspace*{0.5cm}
\includegraphics[width=4.5cm]{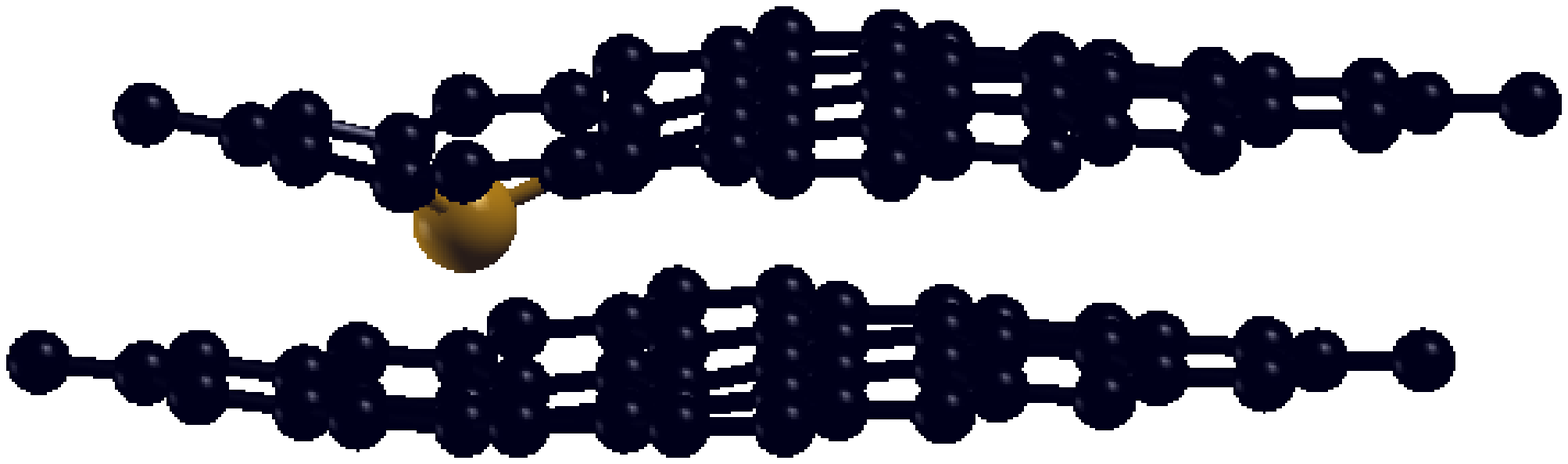} \hspace*{0.5cm}
\includegraphics[width=4.5cm]{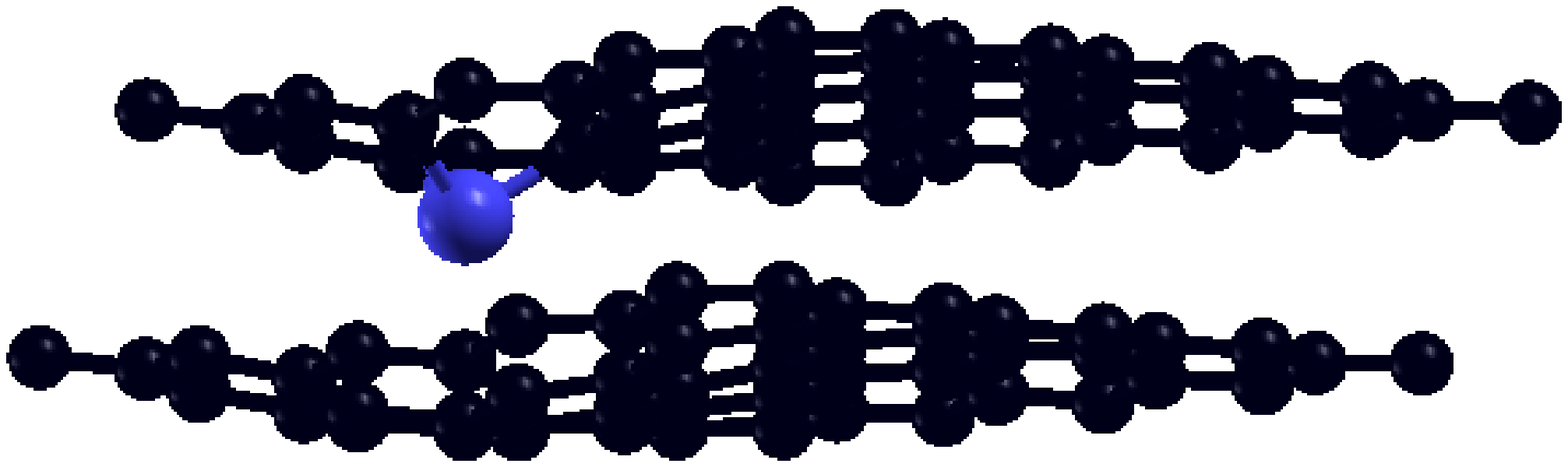} \\
\caption{BLG structures doped with one impurity in the upper layer in $B_1$ position: 1st row -- B, C, N; 2nd row -- Al, Si, P; 3rd row -- Ga, Ge, As.}
\label{struct-1imp}
\end{figure*}

\begin{figure*}[h]
\centering
\includegraphics[width=4.5cm]{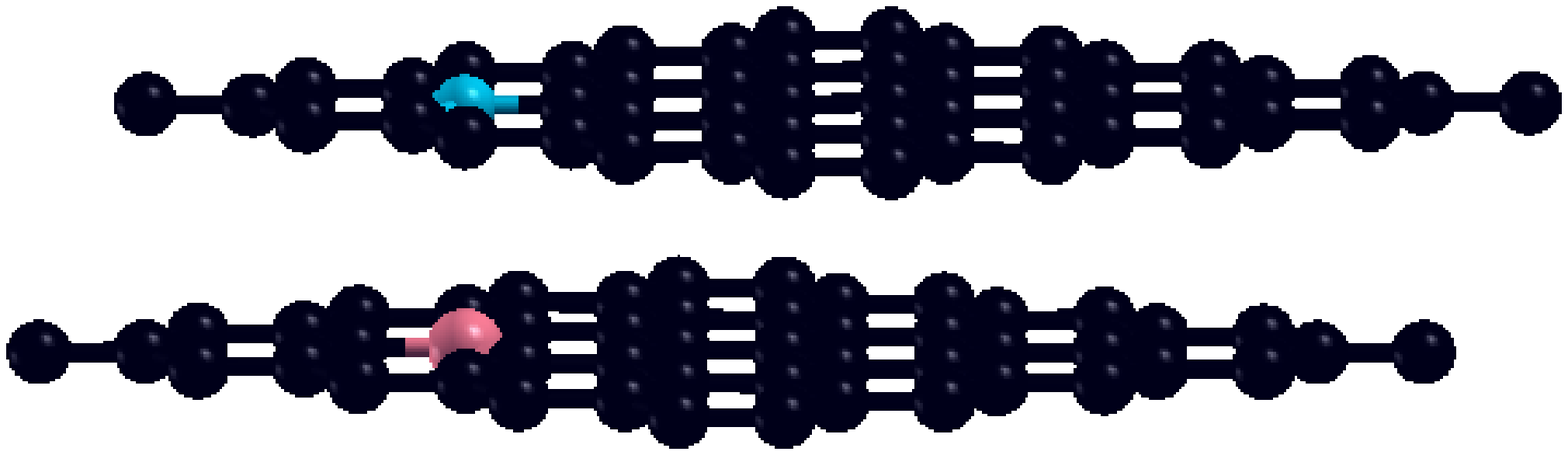} \hspace*{0.5cm} 
\includegraphics[width=4.5cm]{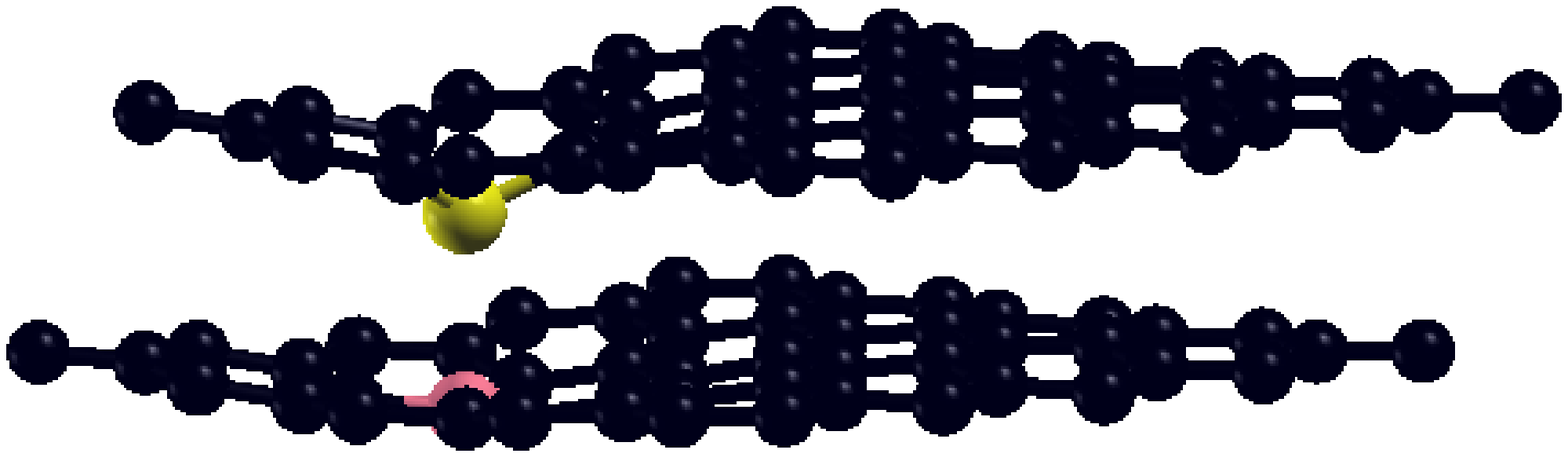} \hspace*{0.5cm}
\includegraphics[width=4.5cm]{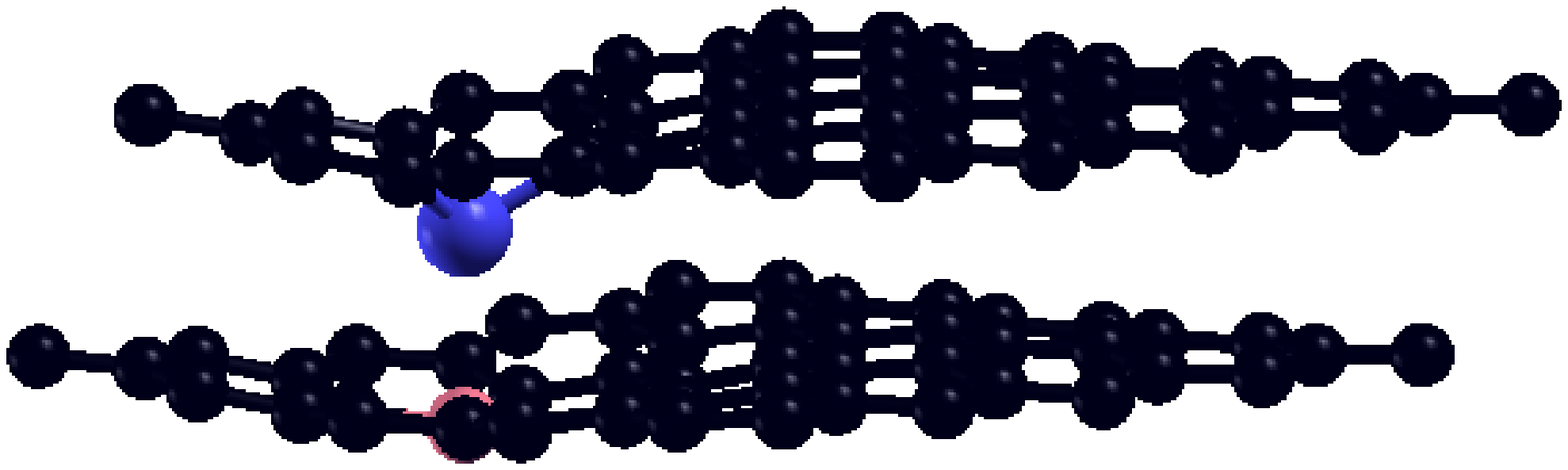} \vspace*{1cm}\\
\includegraphics[width=4.5cm]{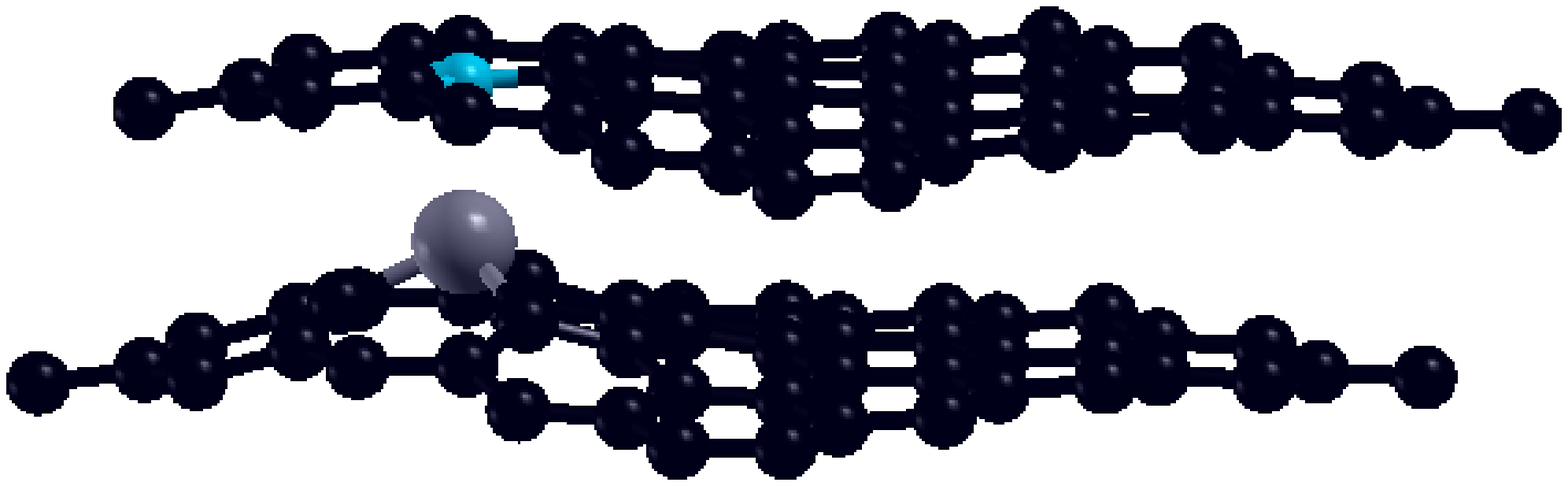} \hspace*{0.5cm}
\includegraphics[width=4.5cm]{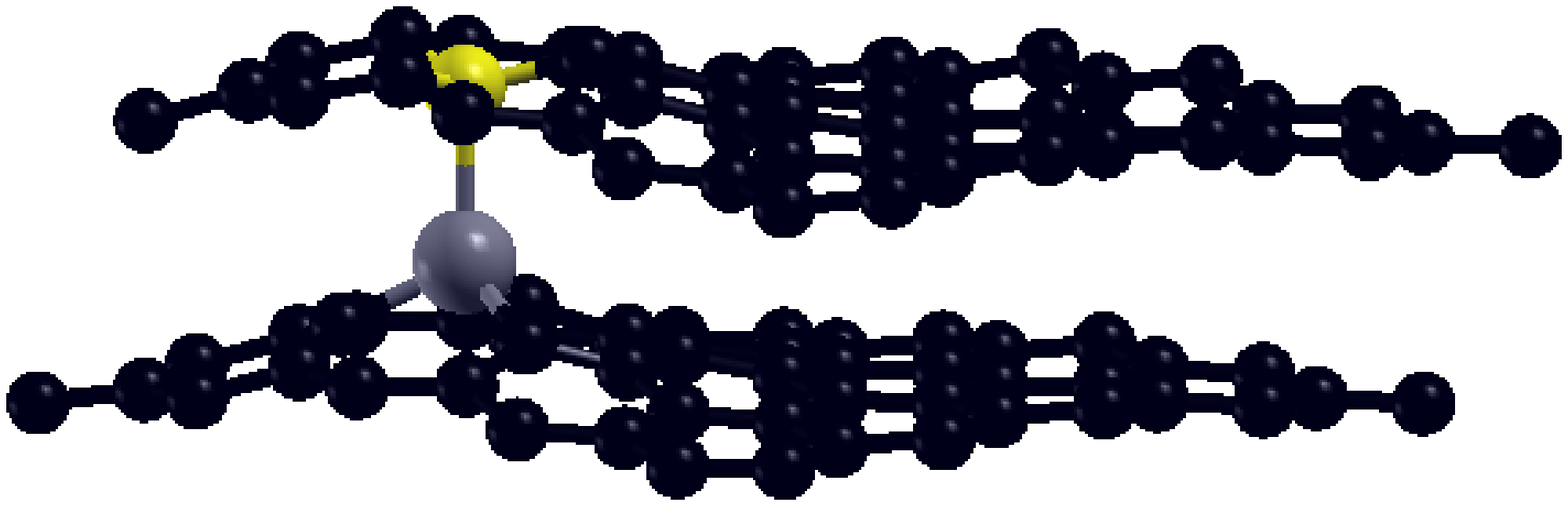} \hspace*{0.5cm}
\includegraphics[width=4.5cm]{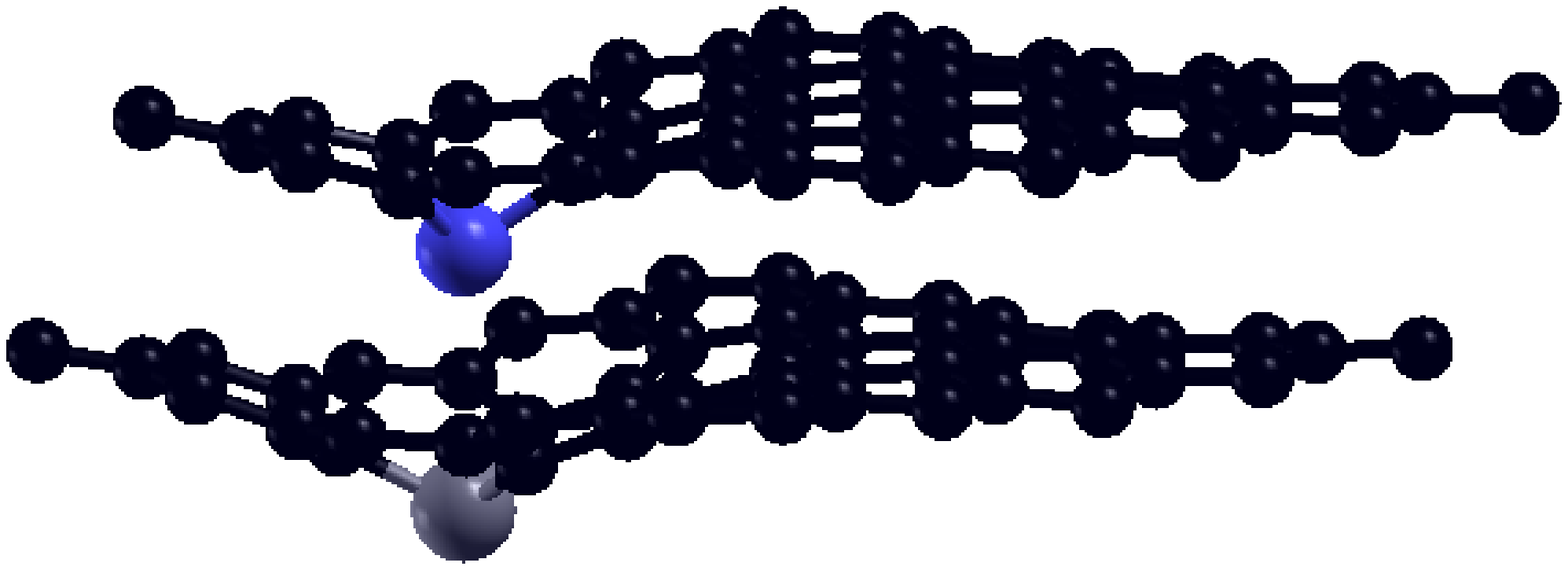} \vspace*{1cm}\\
\includegraphics[width=4.5cm]{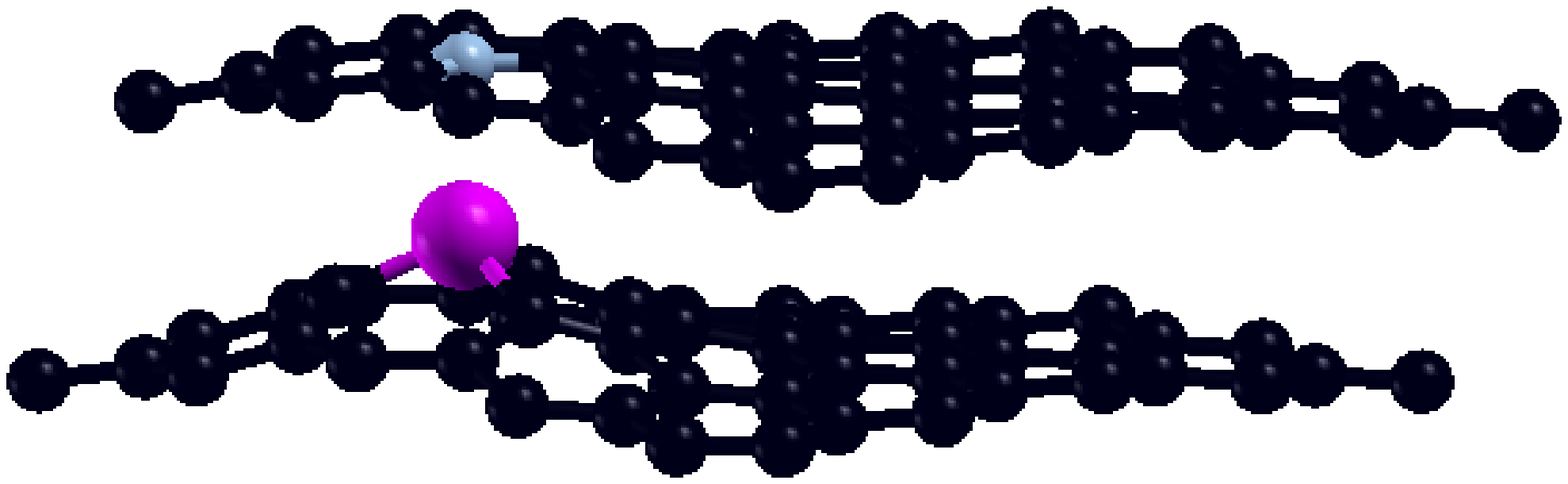} \hspace*{0.5cm}
\includegraphics[width=4.5cm]{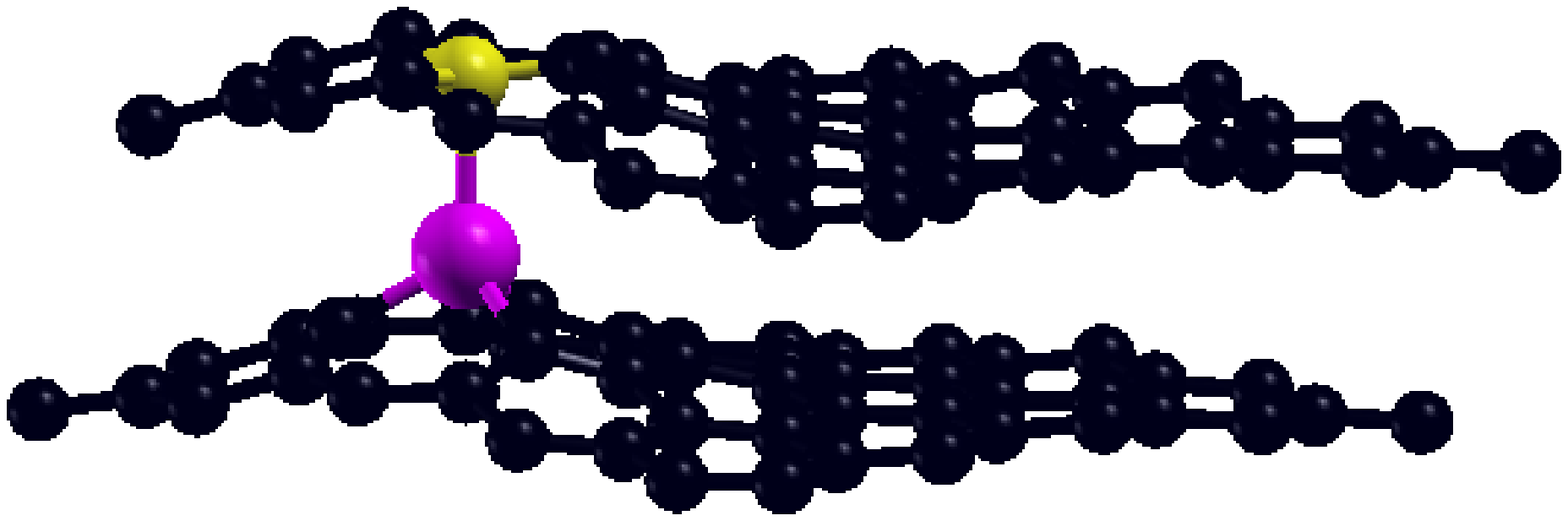} \hspace*{0.5cm}
\includegraphics[width=4.5cm]{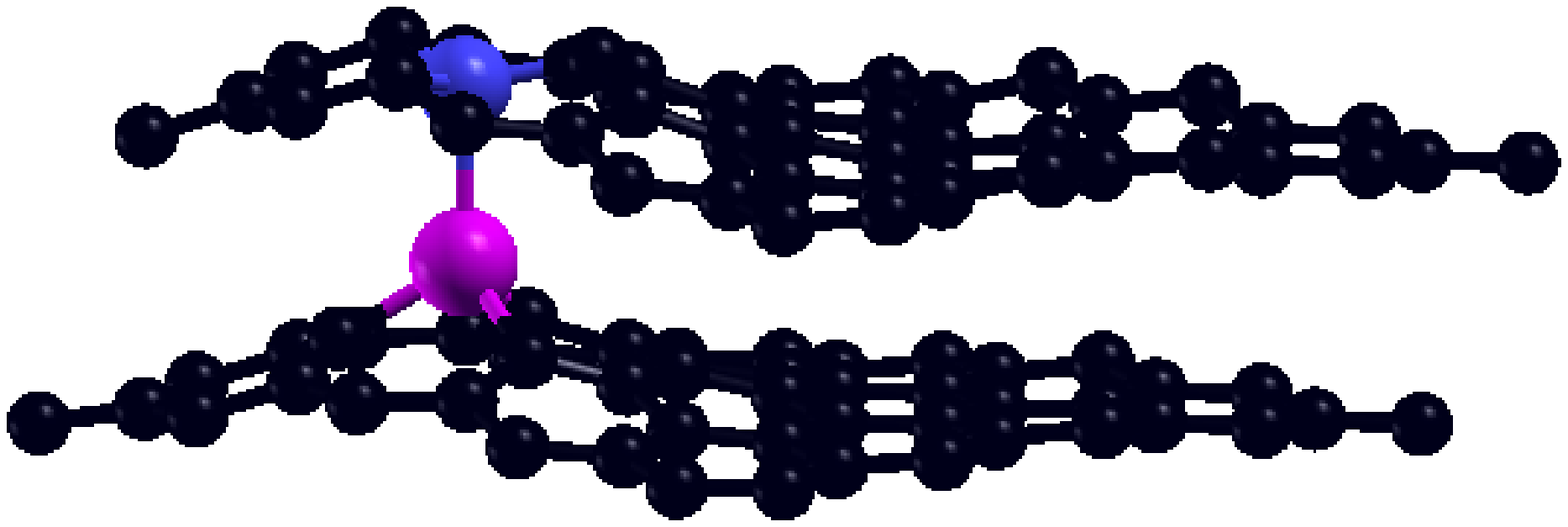} \\
\caption{Dual doped BLG structures, with impurities in $B_1$-$B_2$ positions: 1st row -- B-N, B-P, B-As; 2nd row -- Al-N, Al-P, Al-As; 3rd row -- Ga-N, Ga-P, Ga-As.}
\label{struct-2imps}
\end{figure*}


\begin{table}[h]
\centering
\caption{Structural properties of BLGs with substitutional impurities from  groups III, IV and V in $A_1$ position: average bond length to neighboring C atoms $d_{X-C}$, off-plane shift $\Delta z_X$, BLG inter-layer distance $d_{\rm BLG}$. All distances are given in \AA.}
\vspace*{0.3cm}
\begin{tabular}{c|ccc|cc|ccc}
\hline\hline
d [\AA]              & B    & Al   & Ga   & Si   & Ge   & N    & P    & As  \\
\hline
$d_{X-C}$      & 1.47 & 1.86 & 1.93 & 1.74 & 1.81 & 1.41 & 1.77 & 1.88 \\
$\Delta z_X$  & 0.06 & 1.95 & 2.05 & 1.50 & 1.66 & 0.06 & 1.50 & 1.71 \\
$d_{\rm BLG}$ & 3.36 & 3.58 & 3.59 & 3.53 & 3.56 & 3.36 & 3.53 & 3.57 \\
\hline
\hline
\end{tabular}
\label{tabS1}
\end{table}

\begin{figure*}[h]
\centering
\includegraphics[width=12cm]{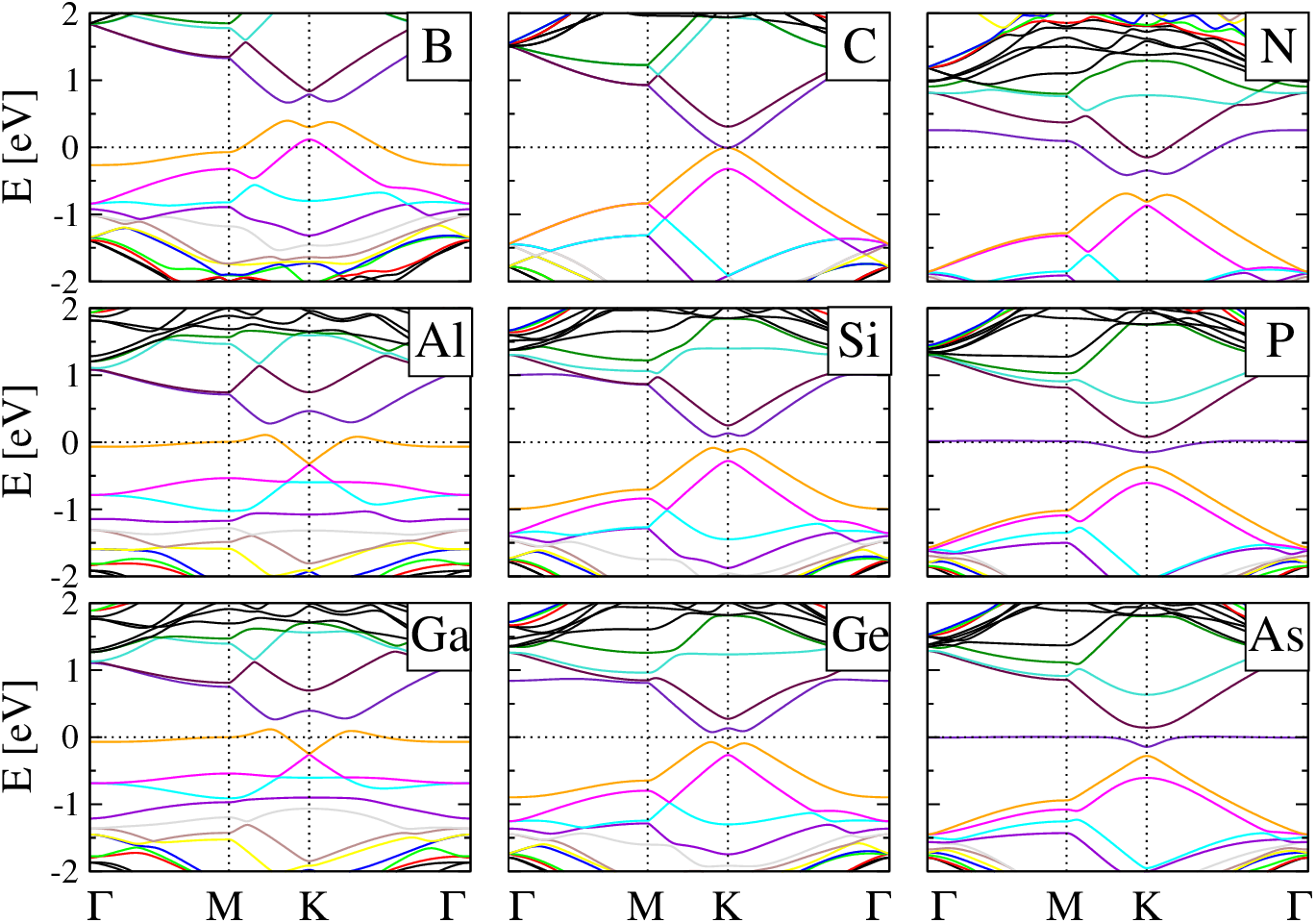}
\caption{Electronic band structures of doped $5\times5$ BLGs with one impurity in $A_1$ position for zero electric field. The impurities correspond to group III (B, Al, Ga), group IV (Si, Ge) and group V (N, P, As) elements. The Fermi level corresponds to $E=0$ eV.}
\label{bands-1imp-poz0}
\end{figure*}

\begin{figure*}[h]
\centering
\includegraphics[width=12cm]{S4}
\caption{The density of states for the doped BLG systems indicated in Fig.\ \ref{bands-1imp-poz0}. In each plot there are three values for the electric field: $E_{\rm field} =$ -0.5 V/nm (blue), 0 V/nm (black) and 0.5 V/nm (red). The Fermi level is marked by vertical dashed lines.}
\label{DOS-1imp-poz0}
\end{figure*}

\begin{figure*}[h]
\centering
\includegraphics[width=12cm]{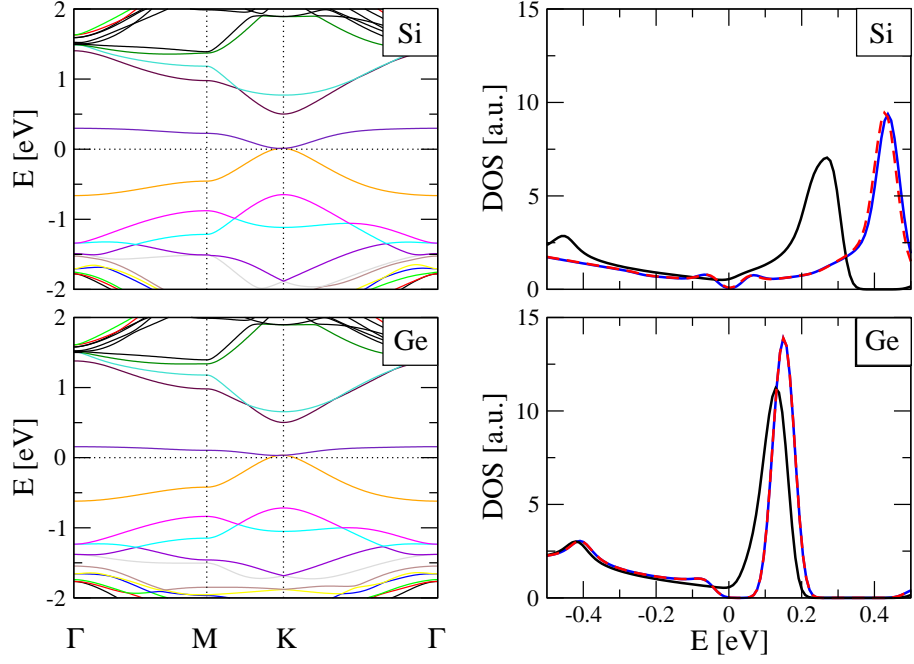}
\caption{Band structures and DOS of BLGs with Si-Si and Ge-Ge substitutions in $B_1$-$B_2$ positions. The systems are gap-less at zero electric field (black). Due to the symmetry, both field orientations are equivalent and a small gap appears for $E_{\rm field} = \pm0.5$ V/nm (red/blue).}
\label{Si-Ge}
\end{figure*}

\newpage

\begin{table*}[h]
\centering
\caption{Formation energies [eV] for BLG systems with one substitutional impurity ($A_1$ and $B_1$ positions).}
\vspace*{0.3cm}
\begin{tabular}{c|ccc|cc|ccc}
\hline\hline
Site  & B    & Al   & Ga  & Si  & Ge  &  N    & P    & As  \\
\hline
$A_1$      &  3.53 & 12.53 & 15.69 &  9.73 & 12.04 &  2.07 &  8.88 & 10.40\\ 
$B_1$      &  3.53 & 12.25 & 15.36 &  9.55 & 11.84 &  2.09 &  8.92 & 10.47\\
\hline
\hline
\end{tabular}
\label{tabS7}
\end{table*}

\vspace*{1cm}

\begin{table*}[h]
\centering
\caption{Thermoelectric behavior at low temperatures of doped BLG systems with impurities in $A_1$ positions, under applied electric field: $\alpha_{\rm CM} = S_{\rm CM} / T $ in units of 10$^{-2}$ $\mu$V/K$^2$.}
\vspace*{0.3cm}
\begin{tabular}{c|ccc|ccc}
\hline\hline
$E_{\rm field}$ [V/nm]  & B    & Al   & Ga   & N    & P    & As  \\
\hline
-5 V/nm      & 6.17 & 20.45 &  18.37 &  -7.94 & -15.44 & -7.76  \\ 
 0           &10.89 &  5.99 &   7.16 &  -7.95 & -42.48 & -20.51  \\
+5 V/nm      & 8.80 &-17.12 & -10.19 &  -5.58 & -84.81 & -45.03  \\
\hline
\hline
\end{tabular}
\label{tabS6}
\end{table*}

\end{document}